\documentclass[11pt]{article}

\usepackage{lineno,hyperref}
\pdfoutput=1
\usepackage{graphics}
\usepackage[russian,english]{babel}
\usepackage[english] {babel}
\usepackage{amssymb,amsfonts,amsmath,mathtext}
\usepackage{graphicx}
\usepackage{subcaption}
\usepackage{longtable}

\usepackage{natbib}
\setcitestyle{authoryear,open={(},close={)}}

\usepackage[nottoc]{tocbibind}

\newcommand{\Ex}{{\mathsf{E}\,}}
\newcommand{\Var}{{\mathsf{Var}}\,}
\newcommand{\R}{{\mathbb R}}

\newcommand{\sd}{{\mathsf{s.d.}}\,}
\newcommand{\rms}{{\mathsf{rms}}\,}

\DeclareMathOperator{\tr}{tr}

\renewcommand{\d}{{\rm d}}
\renewcommand{\Pr}{{\mathsf{P}\,}}
\newcommand{\mean}{{\mathsf{mean}}\,}
\newcommand{\eg}{e.g.\ }

\newcommand{\vs}{vs.\ }
\newcommand{\ie}{i.e.\ }
\newcommand{\wrt}{w.r.t.\ }
\newcommand{\rhs}{r.h.s.\ }

\newcommand{\e}{\rm e}

\modulolinenumbers[5]

   \title{A Hierarchical Bayes Ensemble Kalman Filter
   \footnote{This article is published in Physica D (Nonlinear Phenomena), 2017, v.338, 1-16, \url{doi:10.1016/j.physd.2016.07.009}, free access until January 05, 2017 at \url{https://authors.elsevier.com/a/1U3FZ_3pR42554}.}
   \footnote{This reprint differs from the original article in pagination and typographic detail.}   
   \footnote{\copyright 2016. This manuscript version is made available under the CC-BY-NC-ND 4.0 license http://creativecommons.org/licenses/by-nc-nd/4.0/}
   }
   \author{Michael Tsyrulnikov and Alexander Rakitko\\
   \smallskip
   {\em HydroMetCenter of Russia}\\
   {\small (michael.tsyrulnikov@gmail.com) }}

\begin{document}

\maketitle

\begin{abstract}

A new ensemble filter that allows for the uncertainty in the prior distribution is
proposed and tested. The filter relies on the conditional Gaussian distribution of the state
given the  model-error and predictability-error covariance matrices.
The latter 
are treated as random matrices and updated in a hierarchical Bayes scheme along with the state.
The (hyper)prior distribution of the covariance matrices is assumed to be inverse Wishart.
The new Hierarchical Bayes Ensemble Filter (HBEF) assimilates ensemble members
as generalized observations and allows ordinary observations to influence the covariances. 
The actual probability distribution of the ensemble members is allowed to be different from the true one.
An approximation that leads to a practicable analysis algorithm is proposed. The new filter is studied in numerical experiments with a doubly stochastic one-variable model of  ``truth''. 
The model permits the assessment of the variance of the truth and  
the true filtering error variance at each time instance. 
The HBEF is shown to outperform  the EnKF and the HEnKF by Myrseth and Omre (2010) 
in a wide range of filtering regimes
in terms of  performance of its primary and  secondary filters.

\end{abstract}

\section {Introduction}

Stochastic filtering and smoothing is a mathematical name for what is called in natural sciences data assimilation.
Whenever we have three things: 
(1) an evolving system whose state is of interest to us,
(2) an imperfect mathematical model of the system, and
(3) incomplete and noise-contaminated observations, 
there is room for data assimilation.
Currently, data assimilation techniques are extensively used in geophysics:
meteorology, atmospheric chemistry, oceanography, land hydrology 
\citep[e.g.][]{Lahoz},
underground oil reservoir modeling \cite{Oliver}, 
biogeochemistry \cite{Trudinger},
geomagnetism \cite{Fournier}, 
and being explored in other areas like 
systems biology \cite{yoshida},
epidemiology \cite{rhodes}, 
ecology \cite{niu}, and 
biophysics \cite{Chapelle}.
Data assimilation techniques have reached their most advanced level in meteorology.

To simplify the presentation of our technique, we  confine ourselves 
to sequential discrete-time  filtering, whose goal is to estimate the current state of the 
system given all present and past observations.
This is a cycled procedure, each cycle consists of an 
observation update step (called in meteorology {\em analysis}) when
current observations are  assimilated, and a time update (forecast) step that
propagates information on past observations forward in time.

\subsection {Stochastic models of uncertainty}

Virtually all advanced data assimilation methods rely on stochastic modeling
of the underlying uncertainties in observations and in the forecast model.
Historically, the first breakthrough in meteorological data assimilation was 
the introduction of the stochastic model of locally homogeneous and isotropic  random fields
and the least squares estimation approach based on  correlation functions
(optimal interpolation by Eliassen \cite{Eliassen} and Gandin \cite{Gandin}).
The second big advancement was the development of global multivariate 
forecast error {\em covariance models} no longer based on correlation functions 
but relying on more elaborate approaches like spectral and wavelet models,
spatial filters, diffusion equations, etc. 
\cite{Rabier,Fisher,Deckmyn,Purser,Weaver}; 
these (estimated ``off-line'') forecast-error models have been utilized in so-called
{\em variational} data assimilation schemes \citep[e.g.][]{Rabier}.
The third major invention so far was the  
Ensemble Kalman Filter (EnKF) by Evensen \cite{Evensen},
in which the uncertainty of the system state is assumed to be Gaussian and
represented by a Monte Carlo sample (ensemble), so that
static forecast error covariance models are replaced by dynamic 
and flow-dependent ensemble covariances.
The EnKF has then developed into a wide variety of ensemble based techniques
including ensemble-variational hybrids, \eg  \cite{Houtekamer,Buehner2013,lorenc}.

There is another class of non-parametric Monte Carlo based filters called particle filters \citep[e.g.][]{Leeuwen}.
They do not rely on the Gaussian assumption and thus are better suited to tackle highly
nonlinear problems, but the basic underlying idea of representing the unknown 
continuous probability density by a sum of a relatively small number of delta functions  
looks attractive for low-dimensional systems, whereas in high dimensions,  
its applicability remains to be convincingly shown.
We do not consider particle filters in this paper.

In this research, we propose to retain a kind of Gaussianity because a {\em parametric} prior distribution has the advantage of bringing
a lot of regularizing information in the vast areas of state space, where there are
no nearby ensemble members. But we are going to relax the Gaussian assumption replacing it by 
a more general {\em conditionally} Gaussian model.

\subsection {Uncertainty in the forecast error distribution}

In the traditional EnKF, the forecast (background) uncertainty is characterized 
by the forecast error covariance matrix ${\bf B}$,
which is estimated from the forecast ensemble.
The problem is that this estimate cannot be precise, 
especially in high-dimensional applications of the EnKF, where the
affordable ensemble size is much less than
the dimensionality of state space.
So, the forecast uncertainty in the EnKF is largely uncertain by itself \citep{Furrer,Sacher}.
On the practical side, a common remedy here is a kind of regularization 
of the sample covariance matrix
\citep[e.g.][]{Furrer}.
But these techniques (of which the most widely used is covariance localization or tapering) 
are more or less {\em ad hoc} and have side effects, so
a unifying paradigm to optimize the use of ensemble data in  filtering is needed.
On the theoretical side, there is an appropriate way to account for this 
uncertain uncertainty: hierarchical Bayes modeling (\eg\cite{robert}).

\subsection {Hierarchical Bayes estimation}

In the classical non-Bayesian statistical paradigm, the state ${\bf x}$ ({\em parameter} in statistics)
is considered to be non-random being subject
of estimation from random forecast and random observations.
Optimal interpolation is an example.

In the non-hierarchical Bayesian paradigm, 
both observations ${\bf y}$ and 
the state ${\bf x}$ are regarded as random. 
At the first level of the hierarchy, one specifies the observation likelihood $p({\bf y} | {\bf x})$.
As ${\bf x}$ is random, one introduces the second level of the hierarchy,
the probability distribution of ${\bf x}$ that summarizes our knowledge of the state  ${\bf x}$
before current  observations ${\bf y}$ are taken into account, the
{\em prior}  distribution $p({\bf x} | {\boldsymbol\vartheta})$.
Here ${\boldsymbol\vartheta}$ is the non-random vector of parameters
of the prior distribution (called hyperparameters).
So, the non-hierarchical Bayesian modeling paradigm is, essentially, a two-level hierarchy
(${\bf y} | {\bf x}$ and ${\bf x} | {\boldsymbol\vartheta}$).
In the analysis, the prior density  $p({\bf x} | {\boldsymbol\vartheta})$ is updated using 
the observation likelihood $p({\bf y} | {\bf x})$ leading 
to the {\em posterior} density $p({\bf x} | {\bf y})$.
Note that the analysis step in the Kalman Filter can be viewed as an example of the two-level Bayesian hierarchy, in which
the prior $({\bf x | b,B})$ is the Gaussian distribution with the hyperparameter  ${\bf b}$ being
the predicted ensemble mean vector  and the hyperparameter ${\bf B}$ the  predicted ensemble covariance matrix.
Variational assimilation can  be regarded as a similar two-level Bayesian hierarchy with
${\bf b}$ being the deterministic forecast and ${\bf B}$  the pre-specified covariance matrix. 

In the hierarchical Bayesian paradigm, not only observations and the state are random,
the prior distribution is also assumed to be random (uncertain).
Specifically, the hyperparameters  ${\boldsymbol\vartheta}$  
 are assumed to be random variables having their own
(hyper)prior distribution governed by  hyperhyperparameters $\boldsymbol\gamma$.
If $\boldsymbol\gamma$ are non-random, then we have a three-level hierarchy 
(${\bf y} | {\bf x}$, ${\bf x} | {\boldsymbol\vartheta}$, and 
$\boldsymbol\vartheta | \boldsymbol\gamma$). 
The meaningful number of levels in the hierarchy depends on the observability of the higher-level
hyperparameters: a hyperparameter is worth
to be considered as random and subject of update if it is ``reasonably'' observed.
We will rely in this study on a three-level hierarchy with the prior covariances as the random hyperparameter.

Historically, Le and Zidek \cite{lezidek} introduced uncertain covariance matrices 
in the static geostatistical non-ensemble estimation framework known as Kriging.
Berliner \cite{Berliner} proposed to use the hierarchical Bayesian paradigm to account for
uncertainties in parameters of error statistics used in data assimilation.
Within the EnKF paradigm, 
Myrseth and Omre \cite{Myrseth}  added ${\bf b}$ and ${\bf B}$ to the traditional control vector
assuming that  ${\bf B}$ is the inverse Wishart distributed random matrix and the 
distributions  ${\bf b} | {\bf B}$ and $({\bf x | b,B})$ 
are multivariate Gaussian.
Bocquet \cite{Bocquet} took a different path and treated 
${\bf b}$ and ${\bf B}$ as nuisance variables to be integrated out 
rather than updating them as components of the control vector.
His filter (developed further in \cite{Bocquet2012,Bocquet2015})
imposed  prior distributions for random ${\bf b}$ and ${\bf B}$ in order to change 
the Gaussian prior  of the state  ${\bf x}$
to a more realistic continuous mixture of Gaussians.

In this study, we follow the general path of \cite{Myrseth}.
We propose to split ${\bf B}$ into 
the model error covariance matrix ${\bf Q}$   and 
the predictability error covariance matrix ${\bf P}$.
The reason for such splitting is the fundamentally different nature of model  errors
(which are external to the filter) \vs predictability errors (which are internal, \ie determined by the filter).
At the analysis step, following the hierarchical Bayes paradigm, 
we update ${\bf P}$ and ${\bf Q}$ along with the state ${\bf x}$ 
using both observation and ensemble data.
Performance of the new filter is thoroughly tested in numerical experiments with a one-variable model.
Note that the observation error covariance matrix  is assumed to be precisely known in this study.

\section{Background and notation}
\label{sec_bckg}

We start by outlining filtering techniques that have led to our approach, 
indicating those of their aspects that are relevant  for this paper.
Thereby, we introduce the notation; the whole list of main symbols 
can be found in \ref{app_list}.

\subsection{Bayesian filtering}
\label{sec_bayes}

The general Bayesian filtering paradigm assumes that unknown systems states 
${\bf x}_k \in \R^n$ (where $k=0,1,\dots$ denotes the time instance and $n$ the dimension
of the state space) are random, subject to estimation from random observations 
${\bf y}_{1:k} = ({\bf y}_1,\dots,{\bf y}_k)$.
The true system states obey a Markov stochastic evolutionary model 
such that the {\em transition density}
$p({\bf x}_k | {\bf x}_{k-1})$ is available.
Observations are related to the truth through the observation {\em likelihood} $p({\bf y}_k | {\bf x}_{k})$.
The optimal filtering process consists in alternating forecast and analysis steps.
At the forecast step the {\em predictive} density  $p({\bf x}_k | {\bf y}_{1:k-1})$ is computed.
The goal of the analysis step is to compute the {\em filtering} density  $p({\bf x}_k | {\bf y}_{1:k})$.

At the analysis step, the predictive density is regarded as a prior density, which we denote
by the superscript $f$ (from ``forecast''): 
$p^f({\bf x}_k) = p({\bf x}_k | {\bf y}_{1:k-1})$.
The filtering density can similarly be viewed as the posterior density denoted
by the superscript $a$ (from ``analysis''): 
$p^a({\bf x}_k) = p({\bf x}_k | {\bf y}_{1:k})$.

Direct computations of the predictive and filtering densities are feasible only for very low-dimensional problems.
This difficulty can be alleviated if we turn to {\em linear} systems.

\subsection{Linear observed system}
\label{sec_lin}

The evolution of the truth is  governed  by the 
discrete-time linear  stochastic dynamic system: 
\begin {equation}
\label{mdlx}
{\bf x}_{k}={\bf F}_{k} {\bf x}_{k-1} + {\boldsymbol\varepsilon}_k,
\end {equation}
where ${\bf F}_{k}$ the (linear) forecast operator, 
${\boldsymbol\varepsilon}_k \sim {\cal N}(0, {\bf Q}_k)$ the model error, and ${\bf Q}_k$
the model error covariance matrix.
Observations ${\bf y}_{k}$ are related to the state through the observation equation
\begin {equation}
\label{yHx2}
{\bf y}_{k}= {\bf H}_{k} {\bf x}_{k} + {\boldsymbol\eta}_k,
\end {equation}
where ${\bf H}_{k}$ is the (linear) observation operator,
${\boldsymbol\eta}_k \sim {\cal N}(0, {\bf R}_k)$ the observation error, and ${\bf R}_k$
the observation error covariance matrix.

\subsection{Prior and posterior covariance matrices}
\label{sec_cvm_def}

Here we introduce the prior, posterior, and predictability covariance matrices, which will be extensively 
used throughout the paper.
By ${\bf b}_k = \Ex {\bf x}_k | {\bf y}_{1:k-1}$, we denote the mean of the prior distribution and by
\begin {equation}
\label{B}
{\bf B}_k = \Ex [({\bf x}_k - {\bf b}_k) ({\bf x}_k - {\bf b}_k)^\top | {\bf y}_{1:k-1}]
\end {equation}
the prior covariance matrix.
Similarly, ${\bf a}_k = \Ex {\bf x}_k | {\bf y}_{1:k}$
is the posterior mean and
\begin {equation}
\label{A}
{\bf A}_k = \Ex [({\bf x}_k - {\bf a}_k) ({\bf x}_k - {\bf a}_k)^\top | {\bf y}_{1:k}]
\end {equation}
the posterior covariance matrix.
With the linear dynamics defined in Eq.(\ref{mdlx}), ${\bf b}_k$ and ${\bf B}_k$ 
satisfy the equations
\begin {equation}
\label{ba}
{\bf b}_k = \Ex [ {\bf F}_k {\bf x}_{k-1} + {\boldsymbol\varepsilon}_k | {\bf y}_{1:k-1}] =
                  {\bf F}_k \, {\bf a}_{k-1}
\end {equation}
and
\begin {equation}
\label{BA}
{\bf B}_k = \Ex [ ({\bf F}_k ({\bf x}_{k-1} - {\bf a}_{k-1}) + {\boldsymbol\varepsilon}_k) \cdot
                  ({\bf F}_k ({\bf x}_{k-1} - {\bf a}_{k-1}) + {\boldsymbol\varepsilon}_k)^\top | 
                  {\bf y}_{1:k-1}] = {\bf P}_{k}  + {\bf Q}_{k},
\end {equation}
where 
\begin {equation}
\label{P}
{\bf P}_{k} = {\bf F}_{k} {\bf A}_{k-1} {\bf F}_{k}^\top
\end {equation}
is the predictability (error) covariance matrix.

\subsection{Kalman filter}
\label{sec_kf}

For the linear system introduced in section \ref{sec_lin}, 
the mean-square optimal linear filter is the Kalman filter (KF).
Its forecast step is  
\begin {equation}
\label{kf1}
{\bf x}^f_{k} =  {\bf F}_{k} {\bf x}^a_{k-1},
\end {equation}
where, we recall, the superscripts $f$ and $a$ stand for the forecast and analysis filter 
estimates, respectively.
The analysis update  is
\begin {equation}
\label{kf2}
{\bf x}^a_{k} ={\bf x}^f_{k} + {\bf K}_k ({\bf y}_k - {\bf H}_k {\bf x}^f_k),
\end {equation}
where ${\bf K}_k$ is the so-called gain matrix:
\begin {equation}
\label{gain}
{\bf K}_k = {\bf B}_k {\bf H}_k^\top ({\bf H}_k {\bf B}_k {\bf H}_k^\top + {\bf R}_k)^{-1}.
\end {equation}
The posterior covariance matrix is 
\begin {equation}
\label{kf4}
{\bf A}_{k}=({\bf I - K}_k {\bf H}_k) {\bf B}_k.
\end {equation}
Note that Eqs.(\ref{kf1}) and (\ref{kf2}) constitute the so-called {\em primary filter}
\cite{Dee}, in which the estimates of the {\em state}
are updated. The primary filter uses the forecast error covariance matrix ${\bf B}_k$  computed in the 
{\em secondary filter}, which is comprised of Eqs.(\ref{gain}),(\ref{kf4}), (\ref{BA}), and (\ref{P}).

\subsubsection{Remarks}
\label{sec_remarks_kf}

\begin{enumerate}
\item
\label{remarks_kf_1}
The KF's forecast ${\bf x}^f_{k}$ and analysis ${\bf x}^a_{k}$ are exactly the prior mean
${\bf b}_{k}$ and the posterior mean ${\bf a}_{k}$, respectively.
 Therefore the above  prior and posterior covariance matrices ${\bf B}_k$ and ${\bf A}_k$  
have also the meaning of the {\em error covariance} matrices of the filter's forecast and analysis, respectively.

\item
\label{remarks_kf_2}
The KF's secondary filter uses only observation operators and not observations
themselves. As a consequence, the conditional covariance matrices
${\bf B}_k$, ${\bf A}_k$, and ${\bf P}_k$ coincide  with their unconditional counterparts, 
$\underline{\bf B}_k$, $\underline{\bf A}_k$, and $\underline{\bf P}_k$ 
(this fact will be utilized below in section \ref{sec_est_tru_var}).
\item
The KF produces forecast and analysis estimates ${\bf x}^f_{k}$ and ${\bf x}^a_{k}$ 
that are the best in the mean-square sense among all {\em linear} estimates.
The KF estimates become optimal among {\em all} estimates
if the involved error distributions are Gaussian.
For highly non-Gaussian distributions, the KF can be significantly sub-optimal,
so the (near) Gaussianity is implicitly assumed in the KF
(this  holds for the ensemble KF as well).
\end{enumerate}

The KF is still prohibitively expensive 
in high dimensions. 
This motivated the introduction and wide spread in geophysical and other applications of 
its Monte Carlo based approximation, the ensemble KF.

\subsection{Ensemble Kalman filter (EnKF)}
\label{sec_enkf}

As compared with the KF, the
EnKF replaces the most computer-time demanding step of forecasting ${\bf P}_k$ 
(via Eq.(\ref{P})) by its {\em estimation} 
from a (small) forecast ensemble.
Members of this ensemble, ${\bf x}^{fe}_k(i)$ 
(where $fe$ denotes the forecast ensemble, $i=1,\dots, N$, and
$N$ is the ensemble size)
are generated by replacing  the two uncertain quantities in Eq.(\ref{mdlx}), 
${\bf x}_{k-1}$ and
${\boldsymbol\varepsilon}_{k}$,
by their simulated counterparts,
${\bf x}^{ae}_{k-1}(i)$
and 
${\boldsymbol\varepsilon}^e_{k}(i)$, respectively:
\begin {equation}
\label{enkf1}
{\bf x}^{fe}_k(i)= {\bf F}_{k} {\bf x}^{ae}_{k-1}(i) + {\boldsymbol\varepsilon}^e_{k}(i).
\end {equation}
Here the superscript ${ae}$ stands for the analysis ensemble (see below in this subsection)
and the superscript $e$ for a simulated pseudo-random variable.
Then, the sample $\{ {\bf x}^{fe}_k(i) \}_{i=1}^N$ is used
to compute the sample (ensemble) mean 
and the sample covariance matrix ${\bf S}_k$.
The Kalman gain ${\bf K}_k$ is computed following Eq.(\ref{gain}), in which ${\bf B}_k$ is a  
somehow regularized ${\bf S}_k$ (normally, by applying variance inflation and spatial covariance localization,
\citep[e.g.][]{Furrer}).

The analysis ensemble ${\bf X}^{ae}_{k} = \{ {\bf x}^{ae}_{k}(i) \}$ is computed 
either deterministically by transforming the forecast ensemble \citep[e.g.][]{tippett}, 
or stochastically \citep[e.g.][]{Houtekamer}.
In this study, we make use of the stochastic analysis ensemble generation technique,
in which  the observations are perturbed by adding their simulated observation errors
$\boldsymbol\eta^e(i) \sim {\cal N}({\bf 0,R})$ and then
assimilated  using ${\bf x}^{fe}_k(i)$ as the background:
\begin {equation}
\label{Xae}
{\bf x}^{ae}_k(i) = {\bf x}^{fe}_k(i)  + {\bf K}_k ({\bf y}_k + \boldsymbol\eta^e(i) - {\bf H}{\bf x}^{fe}(i)).
\end {equation}
Note that in practical applications, the forecast operator ${\bf F}_{k}$ is allowed to be nonlinear.

\subsection{Methodological problems in the EnKF that can be alleviated using the hierarchical Bayes approach}
\label{sec_enkf_meth_probl}

\begin{enumerate}
\item
\label{list_enkf_probl_exactB}

In most EnKF applications, 
the prior covariance matrix is largely uncertain
due to the insufficient ensemble size,
which is not optimally accounted for.
As a result, the filter's performance degrades.

\item
\label{list_enkf_probl_feedb}

In  the EnKF analysis equations, there is no intrinsic feedback from observations 
to the forecast error covariances. The secondary filter is completely divorced from the  primary one.
This underuses the observational information (because  observation-minus-forecast differences
 do contain  information on forecast-error covariances) and
requires external adaptation or manual tuning of the filter.

\end{enumerate}

\subsection{Hierarchical filters}
\label{sec_hier}

By hierarchical filters, we mean those that aim at explicitly accounting for the
uncertainties in the filter's error distributions using hierarchical Bayesian modeling.

\subsubsection{Hierarchical Ensemble Kalman filter (HEnKF) by Myrseth and Omre \cite{Myrseth}}
\label{sec_henkf}

Myrseth and Omre \cite{Myrseth} were the first who used the Hierarchical Bayes approach to address the uncertainty
in the forecast error covariance matrix within the EnKF.
Here we outline their technique using our notation.
To simplify the comparison of their filter with ours, we assume that the dynamics are linear
and neglect the uncertainty in the prior mean vector ${\bf b}_k$  identifying it with the 
deterministic forecast ${\bf x}^f_k$.
The HEnKF differs from the EnKF in the following respects.
\begin{description}
\item[(i)]
${\bf B}_k$ is assumed to be a random matrix 
with the inverse Wishart 
prior distribution: 
${\bf B}_k \sim {\cal IW}(\theta, {\bf B}^f_k)$,
where $\theta$ is the  scalar sharpness parameter and ${\bf B}^f_k$ the prior mean 
covariance matrix (see our \ref{app_W}). 
${\bf B}^f_k$ is postulated to be equal to the previous-cycle posterior mean covariance matrix.

\item[(ii)]
The forecast ensemble members are assumed to be drawn from the Gaussian distribution
${\cal N}({\bf b}_k, {\bf B}_k)$,
where ${\bf B}_k$ is the true forecast error covariance matrix.

\item[(iii)]
Having the inverse Wishart prior for ${\bf B}_k$ and independent Gaussian ensemble members drawn
from ${\cal N}({\bf b}_k, {\bf B}_k)$ implies that these  ensemble members can be 
used to refine the prior distribution of ${\bf B}_k$. The respective posterior
distribution of ${\bf B}_k$ is again inverse Wishart with the mean ${\bf B}^a_k$ 
equal to a linear combination
of ${\bf B}^f_k$ and the ensemble covariance matrix ${\bf S}_k$
(see  our 
\ref{app_W1}).

\item[(iv)]
In generating the analysis ensemble members ${\bf x}^{ae}_k(i)$, the HEnKF perturbs not only observations (as in the EnKF)
but also simultaneously the ${\bf B}_k$ matrix according to its posterior distribution.
\end{description}
The HEnKF was shown to outperform the EnKF in numerical experiments with 
simple low-order models for small ensemble sizes, as well as
with an intermediate complexity model without model errors for a constant field \cite{Myrseth}.

\subsubsection{EnKF-N ``without intrinsic need for inflation'' by  Bocquet et al. \cite{Bocquet, Bocquet2012,Bocquet2015}}
\label{sec_enkfN}

In the EnKF-N, the prior mean and covariance matrices are assumed to be uncertain nuisance parameters with non-informative Jeffreys
prior probability distributions. 
There is also a variant of the EnKF-N with an informative Normal-Inverse-Wishart prior for $({\bf b,B})$.
With the Gaussian conditional distribution of the truth $({\bf x | b,B})$ and the perfect ensemble,
the unconditional distribution of the truth given the forecast and the ensemble 
is analytically tractable and is proposed to replace, in the  EnKF-N, the traditional Gaussian prior.
The resulting analysis algorithm involves a non-quadratic minimization problem,
which, as the authors argue, can be feasible in high-dimensional problems.

In numerical experiments with low-order models, the EnKF-N without a superimposed inflation was shown to be competitive
with  the EnKF with optimally tuned inflation. 
There were also indications that the EnKF-N can reduce the need in covariance localization.

\subsubsection{Need for further research}
\label{sec_hier_summ}

Returning to the list of the EnKF's problems (section \ref{sec_enkf_meth_probl}),
we note that the HEnKF does address the first problem 
(the uncertainty in ${\bf B}_k$), but it does not address the second one 
(absence of feedback from observations to covariances in the EnKF).
Next, assumption {\bf (ii)} in section \ref{sec_henkf}
is too optimistic, which will be discussed below in section \ref{sec_pe_lik}
when we introduce our filter.
Finally, the HEnKF is going to be very costly in high dimensions because of the need to sample from 
an inverse Wishart distribution. (Myrseth and Omre \cite{Myrseth} note, though, that this 
computationally heavy sampling can be dropped, but, to the authors' knowledge,
this opportunity has not yet been tested.)

The EnKF-N addresses both problems mentioned in section \ref{sec_enkf_meth_probl},
but it relies on the assumption that forecast ensemble members are drawn
from the same distribution as the truth (like the HEnKF relies on its assumption {\bf (ii)}).
As we will argue in section \ref{sec_pe_lik}, this cannot be guaranteed if background error covariances are uncertain.
Besides, the EnKF-N has no memory in the covariances (as it does not explicitly update them).
As we  show below, updating and cycling the covariances can be useful.

Thus, both the HEnKF and the EnKF-N are important first contributions to the area
of hierarchical filtering, but there is a lot of room in this area for further improvements and new approaches.
This study presents one of them.

\section{Hierarchical Bayes Ensemble (Kalman) Filter (HBEF)}
\label{sec_hbef}

\subsection{Setup and idea}
\label{sec_hbef_idea}

We formulate the HBEF for  linear  dynamics and  linear observations, see
Eqs.(\ref{mdlx}) and (\ref{yHx2}).
Observation errors are Gaussian.
Other settings  come, mainly, from the formulation of conditions under
which the EnKF actually works in geophysical applications:
\begin{enumerate}

\item
\label{list_setup_N}
The ensemble size is too small for sample covariance matrices to be accurate estimators.

\item
\label{list_setup_F}
The direct computation of the predictability covariance matrix 
${\bf P}_{k}$ as 
${\bf F}_{k} {\bf A}_{k-1} {\bf F}_{k}^\top$ is unfeasible.

\item
\label{list_setup_Q}
The model error covariance matrix ${\bf Q}_k$ is temporally variable and  explicitly unknown. 


\end{enumerate}

We also hypothesize that

\begin{enumerate}

\setcounter{enumi}{3}

\item
\label{list_setup_me_distr}
Conditionally on ${\bf Q}_k$, the model errors are zero-mean Gaussian:
$\boldsymbol\varepsilon_k | {\bf Q}_k \sim {\cal N}({\bf 0,Q}_k)$.


\item
\label{list_setup_me}
We can draw independent pseudo-random samples from  ${\cal N}({\bf 0,Q}_k)$ with the true ${\bf Q}_k$.

\end{enumerate}
Under these assumptions, the KF theory cannot be applied. In this research, we propose 
a theory and design a filter (the HBEF) that acknowledge 
in a more systematic way than this is done in the EnKF
 that the covariance matrices ${\bf Q}_k$ and ${\bf P}_k$
are substantially uncertain. 
 We regard ${\bf Q}_k$ and ${\bf P}_k$ as  additional (to the state ${\bf x}_k$) random
matrix variate variables to be estimated along with the state.
We represent both the prior and the posterior distributions hierarchically:
\begin {equation}
\label{xPQ_hier}
p({\bf x,P,Q}) = p({\bf P,Q}) \cdot p({\bf x}  | {\bf P,Q})
\end {equation}
and advance in time the two densities in the r.h.s.\ of this equation. 
Thereby the conditional density 
$p({\bf x}  | {\bf P,Q})$
is shown below to remain Gaussian.
This point is central to our approach. 
As for the marginal density $p({\bf P,Q})$, its exact evolution
appears to be unavailable, so  we introduce approximations to  
the prior, postulating it to be static and based on the inverse Wishart distribution at any assimilation cycle.

Actually, not only ${\bf Q}_k$ and  ${\bf P}_k$ are uncertain, 
the prior conditional mean ${\bf b}_k$  is uncertain as well. 
But to simplify the presentation of our approach,
we disregard the uncertainty in ${\bf b}_k$
and assume that ${\bf b}_k={\bf x}^f_k$, where ${\bf x}^f_k$ is the deterministic forecast.
This implies that remark \ref{remarks_kf_1} in section \ref{sec_remarks_kf} applies here,
therefore we will use the terms ``prior'' and ``forecast error'' interchangeably
(and similarly for ``posterior'' \vs ``analysis error'').
 
A notational comment is in order. To avoid confusion of a point {\em estimate} (produced by a filter)
with its {\em true} counterpart, we mark the former with a superscript ($f$ or $a$) or  the tilde.
E.g.\ ${\bf B}^a_k$ is the analysis point estimate of the true prior variance ${\bf B}_k$.

\subsection{Observation and ensemble data to be assimilated}
\label{sec_hbef_data}

The HBEF aims to optimally assimilate not only conventional observations but also
ensemble members. To estimate ${\bf Q}_k$ and ${\bf P}_k$, we 
split the forecast ensemble (computed on the interval between the time instances $k-1$ and $k$)
${\bf X}^{fe}_{k}=({\bf x}^{fe}_{k}(1), \dots, {\bf x}^{fe}_{k}(N))$
into two ensembles.
The first one is the model error  ensemble
${\bf X}^{me}_{k}=({\bf x}^{me}_{k}(1), \dots, {\bf x}^{me}_{k}(N))$,
whose members are pseudo-random draws from the true distribution of the model errors.
The second ensemble is the predictability ensemble
${\bf X}^{pe}_{k}=({\bf x}^{pe}_{k}(1), \dots, {\bf x}^{pe}_{k}(N))$ defined to be 
the result of the application of the forecast operator ${\bf F}_{k}$ to the previous-cycle 
analysis ensemble ${\bf X}^{ae}_{k-1}$. The latter is generated by the filter
to represent the posterior distribution of the truth (see below).

Note that this splitting of the forecast ensemble
 does {\em not} imply that the ensemble size is doubled.
In the course of the traditional forecast ensemble, we suggest  preventing 
model error perturbations from being added to the model fields
while accumulating them in the model error ensemble members.

We denote the combined (observation and ensemble) data at the  time  $k$ as 
${\bf Y}_k=({\bf y}_k, {\bf X}^{me}_k, {\bf X}^{pe}_k)$.
To assimilate these data, we need the respective likelihoods.

\subsection{Observation likelihood}
\label{sec_obs_lik}

The Gaussianity of observation errors implies that the observation likelihood is, by definition,
\begin {equation}
\label{obslik1}
p({\bf y}_k | {\bf x}_k) \propto
        \e^{ -\frac12 ({\bf y}_k - {\bf H}{\bf x}_k)^\top {\bf R}_k^{-1}
                      ({\bf y}_k - {\bf H}{\bf x}_k) }.
\end {equation}
%

\subsection{Model error ensemble likelihood}
\label{sec_me_lik}

From assumption \ref{list_setup_me} (section \ref{sec_hbef_idea}) and \ref{app_W1}, it follows that 
we can write down the likelihood of ${\bf Q}_k$ given the model error  ensemble member ${\bf x}_k^{me}(i)$:
\begin {equation}
\label{melik}
p({\bf x}_k^{me}(i) | {\bf Q}_k) \propto
        \frac{1} {|{\bf Q}_k|^{1/2}}
        \e^{ -\frac12 ({\bf x}_k^{me}(i))^\top {\bf Q}_k^{-1} {\bf x}_k^{me}(i) },
\end {equation}
where $|.|$ stands for the matrix determinant.

We emphasize that the existence of the likelihood $p({\bf x}_k^{me}(i) |  {\bf Q}_k)$, Eq.(\ref{melik}),
implies that members of the model error ensemble ${\bf X}_k^{me}$ {\bf can be viewed as observations
on the true ${\bf Q}_k$}.
This is because the likelihood provides the necessary relationship 
between the data we have (${\bf x}_k^{me}(i)$ here) and the parameter we aim to estimate
(${\bf Q}_k$), 
see also 
\ref{app_W1}.
For the whole ensemble, the likelihood becomes
\begin {equation}
\label{melik2}
p({\bf X}_k^{me}  |  {\bf Q}_k) = 
  \prod_{i=1}^N p({\bf x}_k^{me}(i) | {\bf Q}_k) \propto  |{\bf Q}|^{-\frac{N}{2}} \, 
   \e^{ -\frac{N}{2} \tr ( {\bf S}_k^{me}  {\bf Q}_k^{-1} ) },
\end {equation}
where 
\begin {equation}
\label{Sme}
{\bf S}_k^{me} = \frac{1}{N} \sum_{i=1}^N {\bf x}_k^{me}(i) \, {\bf x}_k^{me}(i)^\top
\end {equation}
is the sample covariance matrix.

\underline{Remark}.  Equation (\ref{Sme}) differs from the conventional sample covariance formula: 
the ensemble members are not centered by the ensemble mean
 and  the sum is divided by $N$ and not by $N-1$. 
 These differences stem from our neglect of the uncertainty in ${\bf b}_k$. 
In practical problems, when we are not so sure about the mean, the conventional sample
covariance matrix is to be preferred.

\subsection{Predictability  ensemble likelihood}
\label{sec_pe_lik}

Note that both ordinary observations ${\bf y}_k$ and 
model error ensemble members ${\bf x}_{k}^{me}(i)$ are produced outside the filter.
The likelihoods Eqs.(\ref{obslik1}) and (\ref{melik}) relate ${\bf y}_k$ and
${\bf x}_{k}^{me}(i)$ to the variables (${\bf x}_k$ and ${\bf Q}_k$, respectively), which
are independent of the filter, too.
So, the two likelihoods do influence the filter (they are, in fact, parts of 
its setup) but not vice versa.

This is in contrast to the predictability ensemble members ${\bf x}_k^{pe}(i)$, 
which are generated by the filter itself.
For each $k$, both the distribution of ${\bf x}_k^{pe}(i)$
and the true ${\bf P}_k$ are  {\em determined} by the filter's performance.
Therefore, we cannot {\em impose} a relationship between ${\bf x}_k^{pe}(i)$
and ${\bf P}_k$. 
We can only try to {\em reveal} this relationship.

In so doing, we note that the {\em true} prior covariances are unavailable to the filter (assumption \ref{list_setup_N}). Therefore, the analysis gain matrix ${\bf K}_k$ is inevitably inexact
\cite{Furrer,Sacher}, which
causes the analysis ensemble members ${\bf x}_k^{ae}(i)$ to be distributed with a  
covariance matrix different from the true posterior covariance matrix ${\bf A}_{k}$.
 As a result, the  next-cycle
predictability ensemble members ${\bf x}_{k+1}^{pe}(i)$ cannot 
be distributed with the true predictability covariance matrix ${\bf P}_{k+1}$.
(For the same reason, members of
the traditional forecast ensemble ${\bf X}_{k}^{fe}$ cannot have the same conditional distribution as
the truth in any situation in which  ${\bf B}_{k}$ is uncertain.)
This important point is further illustrated below in  sections \ref{sec_Pobs} and \ref{sec_feedb}.

The conclusion that there is no known relationship between ${\bf x}_k^{pe}(i)$
and ${\bf P}_k$ entails that  the likelihood $p({\bf x}_k^{pe}(i) | {\bf P}_k)$
is not available and so, strictly speaking, the predictability ensemble members 
cannot be used (assimilated) to update the prior distribution and
yield the desired posterior distribution of ${\bf P}_k$.
In order to come up with a mathematically sound way of extracting information on the 
true ${\bf P}_k$ contained in the predictability ensemble ${\bf X}_k^{pe}$,
we use the following device.

First, we postulate the existence of an (explicitly unknown) 
auxiliary  matrix variate random variable $\boldsymbol\Pi_{k}$
such that the predictability ensemble members 
${\bf x}_k^{pe}(i)$ are Gaussian distributed with the known 
mean (identified with the deterministic forecast ${\bf x}_k^f$)
and the covariance matrix $\boldsymbol\Pi_{k}$:
\begin {equation}
\label{pilik}
p({\bf X}^{pe}_k  | \boldsymbol\Pi_k) = 
  \prod_{i=1}^N p({\bf x}^{pe}(i) | \boldsymbol\Pi_k) \propto  |\boldsymbol\Pi_k|^{-\frac{N}{2}} \, 
   \e^{ -\frac{N}{2} \tr ( {\bf S}_k^{pe}  \boldsymbol\Pi_k^{-1} ) },
\end {equation}
where 
${\bf S}^{pe}_k$ is the predictability ensemble sample covariance matrix:
\begin {equation}
\label{Spe}
{\bf S}_k^{pe} = \frac{1}{N} \sum_{i=1}^N ({\bf x}_k^{pe}(i) - {\bf x}_k^f) \, 
                                          ({\bf x}_k^{me}(i) - {\bf x}_k^f)^\top.
\end {equation}
Second, we assume that the true ${\bf P}_k$  has a (known) probability distribution related to $\boldsymbol\Pi_{k}$.
Specifically, we assume that
\begin {equation}
\label{IWP}
{\bf P}_k | \boldsymbol\Pi_k \sim {\cal IW}(\theta, \boldsymbol\Pi_k), 
\end {equation}
where  $\theta$ is the sharpness parameter
(see  
\ref{app_W}), which controls the spread of the distribution of ${\bf P}_k$
around its mean  $\boldsymbol\Pi_k$ (the greater  $\theta$ the smaller the spread).

Now, we observe that we have related ${\bf X}_k^{pe}$ to $\boldsymbol\Pi_{k}$
through the density
$p({\bf X}_k^{pe} | \boldsymbol\Pi_{k})$, see Eq.(\ref{pilik}),
and $\boldsymbol\Pi_{k}$ to ${\bf P}_k$
through the density
$p({\bf P}_k | \boldsymbol\Pi_k)$, see Eq.(\ref{IWP}). 
The resulting  indirect relationship between 
${\bf X}_k^{pe}$ and ${\bf P}_k$ will allow us to  
assimilate the former in order to update the latter.

Thus, we have the likelihoods for both  ordinary observations and ensemble data.
Next, we need the prior distribution.

\subsection{Analysis: prior distribution}
\label{sec_hbef_prior}

The analysis control vector comprises ${\bf x}$, ${\bf P}$, and ${\bf Q}$; we also have the auxiliary
variable $\boldsymbol\Pi$ (a nuisance parameter). 
Note that here and elsewhere we drop the time index $k$ whenever all variables in a given equation
pertain to the same assimilation cycle $k$.
We have to define a prior distribution (recall, denoted by 
the superscript $f$) for all these
four variables combined. By the prior distribution, we mean the conditional distribution given all past 
assimilated data ${\bf Y}_{1:k-1}$. This conditioning is implicit throughout the paper 
in pdfs marked by the  superscript $f$. 
We specify the joint prior hierarchically: 
\begin {equation}
\label{pfxPiPQ}
p^f({\bf x}, \boldsymbol\Pi, {\bf P}, {\bf Q}) = p^f( \boldsymbol\Pi, {\bf P}, {\bf Q}) \, 
         p^f({\bf x} |  \boldsymbol\Pi, {\bf P}, {\bf Q}) =
         p^f({\bf Q}) \,
         p^f(\boldsymbol\Pi | {\bf Q}) \,
         p^f({\bf P} | {\bf Q}, \boldsymbol\Pi) \,
         p  ({\bf x} | {\bf P}, {\bf Q}).
\end {equation}
The key feature here (assumed at the start of filtering, \ie at $k=1$, and proved below for $k >1$) 
is that the prior distribution of the state is 
{\em conditionally Gaussian} given  ${\bf P, Q}$:
\begin {equation}
\label{pfx}
{\bf x} | {\bf P}, {\bf Q} \sim {\cal N}({\bf x}^f, {\bf B=P}+{\bf Q}).
\end {equation}
Now, consider the priors for the covariance matrices in Eq.(\ref{pfxPiPQ}).
Starting with $p^f({\bf Q})$, we hypothesize that 
there is a  {\em sufficient statistic} ${\bf Q}^f$ and this sufficient statistic  is
produced by the secondary filter as an estimate of ${\bf Q}$ from past data,
see section \ref{sec_fcst_scnd}.
Then, from sufficiency, the dependency on the past data in 
$p^f({\bf Q}_k) \equiv p({\bf Q}_k | {\bf Y}_{1:k-1})$
can be replaced by the dependency on ${\bf Q}^f$, so that
$p^f({\bf Q}) = p({\bf Q} | {\bf Q}^f)$.
Similarly, we postulate that 
$p^f(\boldsymbol\Pi | {\bf Q})=p(\boldsymbol\Pi | {\bf P}^f)$, where
${\bf P}^f$ is also provided by the secondary filter, and that
$p^f({\bf P} | {\bf Q}, \boldsymbol\Pi) = p({\bf P} | \boldsymbol\Pi)$,
where the latter density is defined in Eq.(\ref{IWP}).
As a result, Eq.(\ref{pfxPiPQ}) writes
\begin {equation}
\label{prior}
p^f({\bf x}, \boldsymbol\Pi, {\bf P}, {\bf Q}) =
         p({\bf Q} | {\bf Q}^f) \,
         p(\boldsymbol\Pi | {\bf P}^f) \,
         p({\bf P} | \boldsymbol\Pi) \,
         p  ({\bf x} | {\bf B= P + Q}).
\end {equation}
Further,  we model 
$p({\bf Q} | {\bf Q}^f)$ and $p(\boldsymbol\Pi | {\bf P}^f)$
using the inverse Wishart distribution:
\begin {equation}
\label{IWQPi}
{\bf Q} | {\bf Q}^f \sim {\cal IW}(\chi, {\bf Q}^f) \qquad \mbox{and} \qquad
\boldsymbol\Pi | {\bf P}^f \sim {\cal IW}(\phi, {\bf P}^f),
\end {equation}
where $\chi$ and $\phi$ are the static sharpness parameters.

To summarize, the prior distribution is given in Eq.(\ref{prior}), where the first three densities in the \rhs
are inverse Wishart and the last one is Gaussian.
Prior to the analysis, we have  
the deterministic forecast ${\bf x}^f$ and
the five parameters of the three (hyper)prior (inverse Wishart) distributions:
${\bf Q}^f$, ${\bf P}^f$, $\chi$, $\phi$, and $\theta$.
Now, we have to update the prior distribution using both ordinary and ensemble observations
and come up with the posterior distribution.

\subsubsection{Remarks}
\label{sec_remarks_prior}

\begin{enumerate}
\item
The conditional Gaussianity is a natural extension of the Gaussian assumption 
made in the KF and the EnKF 
and is crucial to the HBEF as it enables a computationally affordable analysis algorithm.
\item
The choice of the inverse Wishart distribution is motivated by its {\em conjugacy} for the Gaussian likelihood
\cite{AndersonT, Gelman}. Conjugacy means that the posterior pdf belongs to the same
distributional family as the prior. In our case,  the inverse Wishart prior is not fully conjugate
but it greatly simplifies derivations and makes the analysis equations partly analytically tractable.
\end{enumerate}

\subsection{Posterior}
\label{sec_post}

Multiplying the prior  Eq.(\ref{prior}) by the three likelihoods, 
Eqs.(\ref{obslik1}), (\ref{melik2}), and (\ref{pilik}), we obtain the posterior
for the extended control vector $({\bf x,P,Q}, \boldsymbol\Pi)$:
\begin {multline}
\label{post1}
p^a({\bf x,P,Q}, \boldsymbol\Pi) = p^f({\bf x,P,Q}, \boldsymbol\Pi \, | \, {\bf X}^{me}, {\bf X}^{pe},   {\bf y}) \propto \\
         p^f({\bf x,P,Q}, \boldsymbol\Pi)  \cdot
         p({\bf y} | {\bf x})   \cdot
         p({\bf X}^{me}  | {\bf Q})  \cdot
         p({\bf X}^{pe}  |  \boldsymbol\Pi) = \\
     [ p({\bf Q} | {\bf Q}^f) \,  p({\bf X}^{me} |   {\bf Q})  ] \cdot
     [ p(\boldsymbol\Pi | {\bf P}^f) \,  p({\bf X}^{pe}  |  \boldsymbol\Pi) ] \cdot 
     [ p({\bf P} |\boldsymbol\Pi) ] \cdot  
       p({\bf x | B=P+Q}) \cdot  p({\bf y} | {\bf x}) 
\end {multline}
Note that in  densities marked by the superscript $a$,  
the dependency on the  past and present  data ${\bf Y}_{1:k}$ is implicit. 
Now, our goal is to transform Eq.(\ref{post1}) and reduce it to the 
required posterior $p^a({\bf x,P,Q})$.

We start by simplifying the expressions in the first two brackets in the third line of Eq.(\ref{post1}).
These are seen to be
the two prior densities,
$p({\bf Q} | {\bf Q}^f)$ and $p(\boldsymbol\Pi | {\bf P}^f)$, 
updated by the respective ensemble data but not yet by ordinary observations.
For this reason, we call them sub-posterior densities and denote by the tilde.
For the  inverse Wishart priors, Eq.(\ref{IWQPi}), and the likelihoods Eqs.(\ref{melik2}) and (\ref{pilik}), 
the sub-posterior distributions are again inverse Wishart (see \ref{app_W1}):
\begin {equation}
\label{tildepQ}
\widetilde p({\bf Q}) = p({\bf Q} | {\bf Q}^f) \,  p({\bf X}^{me} |   {\bf Q})  
                                   \sim  {\cal IW}(\chi+N, {\bf\widetilde Q}),
\end {equation}
\begin {equation}
\label{tildepPi}
\widetilde p(\boldsymbol\Pi) = 
     p(\boldsymbol\Pi | {\bf P}^f) \,  p({\bf X}^{me}  |  \boldsymbol\Pi)  
                                    \sim {\cal IW}(\phi+N, {\bf\widetilde P}),  
\end {equation}
with the mean values
\begin {equation}
\label{tildeQPi}
{\bf\widetilde Q} = \frac {\chi {\bf Q}^f + N {\bf S}^{me}} {\chi +N} \qquad \mbox{and} \qquad
{\bf\widetilde P} = \frac {\phi {\bf P}^f + N {\bf S}^{pe}} {\phi +N}
\end {equation}
Next, we eliminate the nuisance matrix variate parameter $\boldsymbol\Pi$ from the posterior.
The standard procedure in Bayesian statistics is to  integrate $\boldsymbol\Pi$ out.
But in our case we cannot do so analytically, instead we resort to the 
empirical Bayes approach \cite{CarlinLouis} and replace in the posterior, Eq.(\ref{post1}),
$\boldsymbol\Pi$ with its estimate ${\bf\widetilde P}$ 
(the mean of the sub-posterior distribution Eq.(\ref{tildepPi}) 
defined in Eq.(\ref{tildeQPi})).
This allows us to get rid of the second bracket in Eq.(\ref{post1}) 
 (because the expression there does not depend on the control vector $({\bf x,P,Q})$
and no longer depends on $\boldsymbol\Pi$)
and
replace the third bracket by 
\begin {equation}
\label{tildepP}
\widetilde p({\bf P}) = p({\bf P} | \boldsymbol\Pi = {\bf\widetilde P}) 
                     \sim {\cal IW}(\theta, {\bf\widetilde P})
\end {equation}
(see Eq.(\ref{IWP})).
As a result, we arrive at the following equation for the posterior density
\begin {equation}
\label{post2}
p^a({\bf x,P,Q})  \propto \widetilde p({\bf P})  \, \widetilde p({\bf Q}) \, [ p({\bf x | B}) \,  p({\bf y} | {\bf x}) ],
\end {equation}
where ${\bf B=P+Q}$ and all the terms that contain the state ${\bf x}$ are placed inside the bracket.

To reduce the joint posterior Eq.(\ref{post2}) to the marginal 
posterior of ${\bf P,Q}$ times the conditional posterior of ${\bf x}$ given ${\bf P,Q}$
(\ie to represent the posterior hierarchically), we should integrate ${\bf x}$ out of $p^a({\bf x,P,Q})$.
This can be easily done because both ${\bf x}$-dependent terms in the bracket
are proportional to Gaussian pdfs \wrt ${\bf x}$, see Eqs.(\ref{pfx}) and (\ref{obslik1}),
and so is their product.
To analytically integrate $p^a({\bf x,P,Q})$ over ${\bf x}$, we complete the square  in the exponent 
of the $p({\bf x | B}) \,  p({\bf y} | {\bf x})$ expression (technical details are omitted)
and take into account that the integral of a Gaussian pdf equals one, getting
\begin {equation}
\label{Lo}
l({\bf B|y}) = \int_{{\R}^n} p({\bf x | B}) \,  p({\bf y} | {\bf x}) \,\d {\bf x} \propto 
  \frac{ |{\bf A}|^{\frac{1}{2}}} { |{\bf B}|^{\frac{1}{2}} } \cdot
   \e^{ -\frac{1}{2} ({\bf y - H x}^f)^\top  ({\bf H B H^\top + R})^{-1} ({\bf y - H x}^f) },
\end {equation}
where the matrix ${\bf A}$ is defined below in Eq.(\ref{Ba}).
It is worth noting that $l({\bf B|y})$ defined in Eq.(\ref{Lo})
is, essentially, the {\em observation likelihood of the matrix} 
${\bf B}$ defined as $p({\bf y} | {\bf B})$: indeed, 
$p({\bf y} | {\bf B}) = \int p({\bf y} | {\bf x}) \,   p({\bf x} | {\bf B})  \,\d {\bf x}$,
hence the notation $l({\bf B|y})$.

Now we  obtain the final posterior
\begin {equation}
\label{postxPQ}
p^a({\bf x,P,Q}) = p^a({\bf P,Q}) \cdot p^a({\bf x}  | {\bf P,Q}).
\end {equation}
Here, from Eqs.(\ref{post2})  and (\ref{Lo}), 
\begin {equation}
\label{postPQ}
p^a({\bf P,Q}) = \int p^a({\bf x,P,Q}) \,\d {\bf x} \propto  
      \widetilde p({\bf P}) \, \widetilde p({\bf Q}) \,  l({\bf P+Q|y})
\end {equation} 
is the marginal posterior. Further, from Eqs.(\ref{post2})  and (\ref{postPQ}),
\begin {equation}
\label{postxgPQ}
p^a({\bf x}  | {\bf P,Q}) = \frac{p^a({\bf x,P,Q})} {p^a({\bf P,Q})} \propto p({\bf x | B}) \,  p({\bf y} | {\bf x}) 
                           \sim {\cal N}({\bf m}^a({\bf B}), {\bf A}({\bf B})),
\end {equation}
(where, we recall, ${\bf B=P+Q}$) is the conditional posterior.
In Eq.(\ref{postxgPQ}), the proportionality $\propto$ is \wrt ${\bf x}$
(because $p^a({\bf x}  | {\bf P,Q})$ is a probability density of ${\bf x}$),
\begin {equation}
\label{ma}
{\bf m}^a({\bf B}) =  {\bf x}^f + {\bf A} \, {\bf H}^\top {\bf R}^{-1} ({\bf y - H x}^f)
\end {equation}
is the  conditional posterior expectation of ${\bf x}$, and
\begin {equation}
\label{Ba}
{\bf A} = {\bf A}({\bf B})=({\bf B}^{-1} + {\bf H}^\top {\bf R}^{-1} {\bf H})^{-1}
\end {equation}
is the  conditional posterior (analysis error) covariance matrix.

\subsubsection{Remarks}
\label{sec_remarks_posterior}

\begin{enumerate}
\item
\label{remark_Gau_anls}
{ \em Preservation of the conditional Gaussianity in the analysis.}
The posterior conditional distribution of the state  $p^a({\bf x}  | {\bf P,Q})$,  Eq.(\ref{postxgPQ}),
appears to be Gaussian (coinciding with the traditional KF posterior
given ${\bf B=P+Q}$, therefore Eqs.(\ref{ma}) and (\ref{Ba}) are exactly the KF equations).
 So, the conditional Gaussianity ``survives''
the analysis step.

\item
The inverse Wishart priors for the covariance matrices  significantly simplify
the derivation of the posterior distribution, but at the expense of not solving
the problem of noisy long-distance covariances. This implies that covariance
localization should be applied to the ensemble covariances.
\item
The linear combinations of the prior and ensemble covariance matrices
in Eq.(\ref{tildeQPi}) resemble, on the one hand, 
the shrinkage estimator of a covariance matrix proposed by \cite{ledoit} and,
on the other hand, the use of static and ensemble covariances 
in hybrid ensemble variational techniques \citep[e.g.][]{Buehner2013,lorenc}.
\item
Equation (\ref{Lo}) shows that observations do influence the observation likelihood of  ${\bf B}$
(through the innovation vector ${\bf y - H x}^f$), hence they do influence the marginal posterior 
$p^a({\bf P,Q})$, see Eq. (\ref{postPQ}). 
This is the ``mechanism'' in the HBEF that
provides the desired and absent in the KF, EnKF, and HEnKF feedback from observations to the forecast error 
covariances.
\item
In the classical Bayesian filtering theory outlined in section \ref{sec_bayes}, the predictive 
and filtering distributions are conditioned
on {\em ordinary} observations ${\bf y}$. In the HBEF, we explicitly condition the posterior
on both observation and ensemble data  ${\bf Y}$. The two conditionings lead to different
results, but this difference is an inevitable consequence of approximations due to the 
use of the ensemble (Monte Carlo) approach.
We will not distinguish between them in the sequel.

\end{enumerate}

\subsection{Analysis equations}
\label{sec_an}

Having the posterior $p^a({\bf x,P,Q})$, see Eqs.(\ref{postxPQ})--(\ref{postxgPQ}),
we  now need equations to compute quantities needed for the next assimilation cycle.
These are, first, point estimates of ${\bf x,P,Q}$ (which we call deterministic analyses) and second, the
analysis  ensemble  ${\bf X}^{ae}$.

\subsubsection{Posterior mean ${\bf x,P,Q}$}
\label{sec_full_an}

The deterministic analyses   ${\bf x}^a, {\bf P}^a, {\bf Q}^a$ are defined as approximations to their respective posterior mean values.
The latter are given, obviously, by the following equations
\begin {equation}
\label{PQaDet}
{\bf P}^a = \Ex {\bf P} = 
    \int \int p^a({\bf P,Q})\, {\bf P} \, \d {\bf P}  \d {\bf Q}, \qquad
    {\bf Q}^a = \Ex {\bf Q} = 
    \int \int p^a({\bf P,Q}) \,{\bf Q} \, \d {\bf P}  \d {\bf Q},
\end {equation}
\begin {equation}
\label{BaDet}
{\bf B}^a = {\bf P}^a + {\bf Q}^a,
\end {equation}
\begin {equation}
\label{xaDet}
{\bf x}^a = \Ex {\bf x} =  \Ex \Ex ({\bf x|P,Q}) = \Ex {\bf m}^a({\bf P+Q}) = 
    \int \int p^a({\bf P,Q})  \, {\bf m}^a({\bf P+Q}) \, \d {\bf P}  \d {\bf Q},
\end {equation}
where ${\bf m}^a({\bf B})$ is given by Eq.(\ref{ma}),
$p^a({\bf P,Q})$ by Eq.(\ref{postPQ}), 
the expectation is over the posterior distribution,
 and the integration \wrt a {\em matrix} is explained in 
\ref{app_mx_intgl}. 
 
The integrals in Eqs.(\ref{PQaDet}) and (\ref{xaDet}) are not analytically tractable, so we introduce
approximations.
We present here two versions of the analysis equations: a Monte Carlo based and an empirical Bayes based
(the  simplest version).

\subsubsection{Monte Carlo based deterministic analysis}
\label{sec_MC_an}

Here, we approximate the integrals in Eqs.(\ref{PQaDet}) and (\ref{xaDet}) 
using Monte Carlo simulation. More specifically, we employ the importance sampling technique \citep[e.g.][]{Kroese}
with the proposal density 
$\widetilde p({\bf P})  \, \widetilde p({\bf Q})$.
Generating the Monte Carlo draws 
${\bf P}^e(i) \sim \widetilde p({\bf P})$, \,
${\bf Q}^e(i) \sim \widetilde p({\bf Q})$ (where $i=1,\dots,M$ and $M$ is the size of the Monte Carlo sample), and
computing ${\bf B}^e(i) = {\bf P}^e(i) + {\bf Q}^e(i)$, we obtain
the estimates:
\begin {equation}
\label{PQaDetMC}
{\bf P}^a = \frac { \sum_{i=1}^M  l[{\bf B}^e(i) | {\bf y}]  \cdot {\bf P}^e(i)}
                           { \sum_{i=1}^M  l[{\bf B}^e(i) | {\bf y}] }, \qquad
          {\bf Q}^a = \frac { \sum_{i=1}^M  l[{\bf B}^e(i) | {\bf y}]  \cdot {\bf Q}^e(i)}
                           { \sum_{i=1}^M  l[{\bf B}^e(i) | {\bf y}] },
\end {equation}
\begin {equation}
\label{xaDetMC}
{\bf x}^a = \frac { \sum_{i=1}^M  l[{\bf B}^e(i) | {\bf y}]  \cdot 
                                              {\bf m}^a[{\bf B}^e(i)]}
                          { \sum_{i=1}^M  l[{\bf B}^e(i) | {\bf y}] }.
\end {equation}
Note that in view of Eq.(\ref{Lo}), the resulting analysis is nonlinear in both ${\bf x}^f$ and ${\bf y}$.

Sampling from an inverse Wishart distribution can be expensive in high dimensions,
so we propose, next, a cheap alternative.

\subsubsection{The simplest deterministic analysis}
\label{sec_simpl_an}

Here, we neglect the $l({\bf B|y})$ term in Eq.(\ref{postPQ}) altogether, thus
allowing, as in the HEnKF, no feedback from observations to the covariances.
The reason for this neglect is that the
information on ${\bf P}$ and ${\bf Q}$
that comes, first, from the prior matrices ${\bf P}^f$ and ${\bf Q}^f$ and second,
from the two ensembles ${\bf X}^{pe}$ and ${\bf X}^{me}$, summarized in the sub-posterior
distributions $\widetilde p({\bf P})$ and $\widetilde p({\bf Q})$, is much richer than  information
on ${\bf P}$ and ${\bf Q}$ that comes from current observations through the  $l({\bf B|y})$ term.
Indeed, ${\bf P}^f$ and ${\bf Q}^f$  accumulate vast amounts of past (albeit aging) 
information on ${\bf P}$ and ${\bf Q}$.
Model error ensemble members constitute, as we have discussed, $N$ direct observations
on ${\bf Q}$.
Predictability ensemble members are $N$ observations
on $\boldsymbol\Pi$ (and so indirectly on ${\bf P}$). 
But there is only one set of current ordinary observations, 
that is, all current observations combined 
give rise to only one (very) noise contaminated observation on ${\bf H B} {\bf H}^\top +{\bf R}$
(but note that with the known ${\bf R}$, this is the only observation on the {\em true} ${\bf B}$).
Therefore, we assume that  in Eq.(\ref{postPQ})
 $\widetilde p({\bf P})$ and $\widetilde p({\bf Q})$ are much more peaked
\wrt $({\bf P},{\bf Q})$
than $l({\bf P+Q | y})$, so that the correction made to the sub-posterior 
by the relatively flat $l({\bf P+Q | y})$ is rather small, and in the first approximation can be disregarded.
This simplification results in the  marginal posterior
\begin {equation}
\label{pQPa1}
p^a({\bf P,Q}) = \widetilde p({\bf P}) \cdot \widetilde p({\bf Q}).
\end {equation}
Both $\widetilde p({\bf P})$ and $\widetilde p({\bf Q})$ are inverse Wishart pdfs with the mean values
${\bf\widetilde P}$ and ${\bf\widetilde Q}$, respectively, so
\begin {equation}
\label{QPa_simplest}
{\bf P}^a = {\bf\widetilde P} \quad \mbox{and} \quad
{\bf Q}^a = {\bf\widetilde Q}.
\end {equation}
As for the deterministic analysis of the state, 
the integral in Eq.(\ref{xaDet}) remains analytically intractable, so we resort 
to the empirical Bayes estimate
\begin {equation}
\label{xaDet2}
{\bf x}^a = {\bf m}^a({\bf B}^a),
\end {equation}
which is just the KF's analysis with ${\bf B}^a = {\bf P}^a + {\bf Q}^a$ as the assumed forecast error
covariance matrix.

\subsubsection{Analysis ensemble}
\label{sec_an_ens}

Here,  the HBEF follows the stochastic EnKF, see Eq.(\ref{Xae}),
where  ${\bf x}^{fe}(i) = {\bf x}^{pe}(i) + {\bf x}^{me}(i)$.

\subsection{Forecast step}
\label{sec_fcst}

\subsubsection{Primary filter}
\label{sec_fcst_prim}

From Eq.(\ref{mdlx}) and assumption \ref{list_setup_me_distr}, we have
\begin {equation}
\label{prim_x}
{\bf x}^f_{k} = \Ex {\bf x}_{k} | {\bf y}_{1:k-1} =  
                      {\bf F}_{k} \cdot \Ex {\bf x}_{k-1} | {\bf y}_{1:k-1} = {\bf F}_{k} {\bf x}^a_{k-1},
\end {equation}
which is essentially the KF's Eq.(\ref{kf1}).

\subsubsection{Preservation of the conditional Gaussianity}
\label{sec_fcst_Gau}

Let us look at the basic state evolution Eq.(\ref{mdlx}). In that equation,
${\boldsymbol\varepsilon}_k | {\bf Q}_{k}$ is Gaussian and independent of ${\bf x}_{k-1}$.
Further, in the posterior at step $k-1$, as it follows from Eqs.(\ref{postxgPQ}) and (\ref{ma}),
${\bf x}_{k-1}$ is conditionally Gaussian given ${\bf A}_{k-1}$.
Therefore, from
${\bf x}_{k} = {\bf F}_{k} {\bf x}_{k-1} + {\boldsymbol\varepsilon}_k$,
we obtain that ${\bf x}_{k} | {\bf A}_{k-1}, {\bf Q}_{k}$ is Gaussian.
But if we examine the distribution in question ${\bf x}_{k} | {\bf P}_{k}, {\bf Q}_{k}$, 
we observe that with the additional technical assumption that  ${\bf F}_{k}$ is invertible,
conditioning on ${\bf P}_{k}$ is equivalent to conditioning on ${\bf A}_{k-1}$
(in view of Eq.(\ref{P})).
Consequently, Gaussianity of 
${\bf x}_{k} | {\bf A}_{k-1}, {\bf Q}_{k}$ 
implies Gaussianity of 
${\bf x}_{k} | {\bf P}_{k}, {\bf Q}_{k}$.

Thus, the basic HBEF's conditional Gaussianity assumption is preserved at the forecast step
(as well as the analysis step, see remark \ref{remark_Gau_anls} in section \ref{sec_remarks_posterior}).

\subsubsection{Secondary filter}
\label{sec_fcst_scnd}

At the forecast step, the secondary filter has to produce  
${\bf P}^f$ and ${\bf Q}^f$ at the next assimilation cycle.
We postulate persistence as the simplest evolution model for both ${\bf P}$ and ${\bf Q}$, so that
\begin {equation}
\label{fcPQ}
{\bf P}^f_k = {\bf P}^a_{k-1} \qquad \mbox{and} \qquad {\bf Q}^f_{k}={\bf Q}^a_{k-1}.
\end {equation}
%

\subsubsection{Generation of the forecast ensembles}

The predictability ensemble  ${\bf X}^{pe}_k$ is generated by simply applying the forecast operator 
${\bf F}_k$ to the analysis ensemble members ${\bf x}_{k-1}^{ae}(i)$, see section \ref{sec_an_ens}.
The model error ensemble  ${\bf X}^{me}_k$ is generated by directly sampling from the model error distribution ${\cal N}({\bf 0, Q}_k)$.

\section {Numerical experiments with a one-variable model}
\label{sec_expm}

In this proof-of-concept study, we tested the proposed filtering methodology in numerical experiments with
a  one-variable model of ``truth'', so that we were able to draw justified conclusions on 
fundamental aspects of the HBEF.
Note that in the case of the one-dimensional state space we follow the default multi-dimensional 
notation but without the bold face.

We compared the HBEF with 
\begin{enumerate}
\item{\em The reference KF}
that has access to  the ``true'' model error variances $Q_k$ 
and is allowed to directly compute $P_k=F_k A_{k-1} F_k^\top =  F_k^2 A_{k-1}$.

\item{\em The stochastic EnKF}
with the optimally tuned variance inflation factor.

\item{\em The Var}, the filter based on the analysis that uses the constant 
${\bar B}$ (the abbreviation Var stands for the variational analysis, which normally uses the
time-mean   ${B}$).

\item{\em The HEnKF}, in which, we recall, the prior mean is excluded from the analysis control vector in order
to make it comparable with the HBEF.

\end{enumerate}
We evaluated the performance of each filter by two criteria. 
The first (main) criterion reflects the accuracy of the {\em primary filter}
measured by the root-mean-square error (RMSE) of the filter's deterministic analysis of the state.
For any filter except the Monte Carlo based HBEF,
the deterministic analysis is defined to be the standard KF's analysis computed with the deterministic
forecast as the background and the forecast error covariance matrix provided by the 
respective filter. (With the forecast model described below and small ensemble sizes,
the deterministic forecast appeared to work better than the ensemble mean.)

The second criterion represents the accuracy of the {\em secondary filter} in terms of 
the RMSE  of the filter's  estimates of $B$ (for details, see section \ref{sec_ver_scnd} below). 
Note that by the RMSE we understand the root-mean-square difference with the truth (the true $B$ 
is defined below in section \ref{sec_est_tru_var}).

Besides the formal evaluation of the performance of the new filter, we also examined some other
important  aspects of the 
technique proposed. First, we verified that the conditional  distribution of the state given the covariances 
was indeed Gaussian. Second, we confirmed that the forecast ensemble variances
were often systematically different from the true error variances. 
Third, we evaluated the role
of the feedback from observations to the covariances, 
which is present in the HBEF with the Monte Carlo based analysis and absent in the other filters.

To conduct the numerical experiments presented in this paper, we developed a software package
in the R language. The code, which allows one to reproduce all  the below experiments,
and its description are 
available from \verb"https://github.com/rakitko/hbef".

\subsection{Model of ``truth''}

We wish the time series of the truth to resemble the natural variability of geophysical,
specifically, atmospheric fields like temperature or winds. We would also like to be able
to change various aspects of the probability distribution of our modeling true time series,
so that the model of truth be conveniently parametrized, with parameters controlling
distinct features of the time series distribution.

\subsubsection{Model equations}
\label{sec_truth_mdl}

We start by postulating the basic discrete-time equation
\begin {equation}
\label{emdlx}
x_{k}=F_{k} x_{k-1} + \sigma_{k} \varepsilon_k,
\end {equation}
where $x_{k}$ is the truth, $F_{k}$  and ${\sigma}_{k}$ are the scalars to be specified,
and $\varepsilon_k \sim {\cal N}(0,1)$ is the driving discrete-time white noise.
Given the sequences $\{ F_{k} \}$ and $\{ \sigma_{k} \}$, the solution to Eq.(\ref{emdlx}) is 
a Gaussian distributed non-stationary time series. 
The forecast operator $F_{k}$ determines the time-dependent time scale of $x_k$
or, in other words, controls the degree of stability of the system:
forecast perturbations are amplified if $|F_k|>1$ and damped otherwise. 
Both $\{ F_{k} \}$ and $\{ \sigma_{k} \}$ together determine the time-dependent variance 
$V_k$ of the random process $x_k$. 
The noise multiplier $\sigma_{k}$ is the model error standard deviation: $Q_k=\sigma_{k}^2$.

In nature, both the  variance and the temporal length scale
exhibit significant chaotic day-to-day changes. In order to simulate these changes
(and thus to introduce intermittent non-stationarity in the process $x_{k}$), we let 
$F_{k}$ and $\sigma_{k}$ be random sequences by themselves, thus making our model 
{\em doubly stochastic} \cite{tjostheim}.
Specifically, let $F_{k}$ be governed by the equation:
\begin {equation}
\label{mdlA}
F_{k} - {\bar F}=\mu (F_{k-1} - {\bar F}) + \sigma_F {\varepsilon}^F_k,
\end {equation}
where $\mu \in (0,1)$ is the scalar controlling the temporal length scale of the process $F_k$,
$\sigma_F$ is the scalar controlling, together with $\mu$, the variance of $F_{k}$,
${\varepsilon}^F_k$ is  the driving ${\cal N}(0,1)$ white sequence,
and ${\bar F}$ is the mean level of the $F_k$ process. 
Equation (\ref{mdlA}) is the classical first-order auto-regression and its solution $F_{k}$
is a stationary random process.

Further, let $\sigma_{k}$ (see Eq.(\ref{emdlx}))
be a log-Gaussian distributed (which prevents $\sigma$
from attaining unrealistically close to zero values and makes it positive)
stationary time series:
\begin {equation}
\label{eSigma}
{\sigma}_{k} = \exp ( {\Sigma}_{k} )  \qquad \mbox{with} \qquad
{\Sigma}_{k} =\varkappa {\Sigma}_{k-1}  + {\sigma}_\Sigma {\varepsilon}^\Sigma_k.
\end {equation}
Here, $\varkappa$, ${\sigma}_\Sigma$, and ${\varepsilon}^\Sigma$ have the same meanings as their counterparts in 
Eq.(\ref{mdlA}): 
$\mu$, ${\sigma}_F$, and ${\varepsilon}^F$, respectively.
We finally assume that the three random sources in our model, namely,
${\varepsilon}_k$, ${\varepsilon}^F_k$, and ${\varepsilon}^\Sigma_k$ are
mutually independent.
Note that the process $x_{k}$ is  conditionally, given $\{ F_{k} \}$ and $\{ \sigma_{k} \}$, Gaussian, whereas
unconditionally, the distribution of $x_{k}$ is non-Gaussian.

\subsubsection{Comparison with the existing models of ``truth''}
\label{sec_truth_mdl_compar}

The difference of our model from popular simple nonlinear deterministic models,
\eg the three-variable Lorenz model \cite{Lorenz63}
or discrete-time maps used to test data assimilation techniques  (say, logistic or Henon maps \cite{Smith}),
is that in the deterministic models instabilities are curbed
by the nonlinearity, whereas in our model, these are limited by the time 
the random process $|F_k|$ remains above 1. 
The nonlinear deterministic models are chaotic whereas our model is stochastic.

One advantage of our model of truth is 
that it allows us to know not only the truth itself but also its time-specific variance $V_k$.
Indeed, running the model Eq.(\ref{emdlx})   $L$  times with independent realizations of the forcing process  $\varepsilon_k$
(and with the sequences $F_k$ and $\sigma_k$ fixed),
we can easily assess ${V}_k$ using square averaging of $x_k$ over the $L$ realizations.

Another advantage of the proposed model of truth is that it has as many as five independent parameters,
$\bar F$, $\mu$, $\varkappa$, ${\sigma}_F$, and ${\sigma}_\Sigma$, which can be independently changed 
and which control different 
important features of the stochastic dynamical system Eqs.(\ref{emdlx})--(\ref{eSigma}).
These features include magnitudes  and time scales
of the solution $x_k$, the model error variance $Q_k$, and 
 the degree of stability of the system.
Note that these aspects affect the behavior of not only the truth but also the filters
we are going to test.

In addition, the linearity of our model of truth allows the use of the exact KF
as an unbeatable benchmark, which again would not be possible with nonlinear 
deterministic models of truth.

Finally, we remark that the model defined by Eqs.(\ref{emdlx})--(\ref{eSigma}) is, actually, 
{\em nonlinear} if regarded as a state-space model, \ie if  the model equations are written as
a Markov model for the {\em vector} state variable $(x_k, F_k, \Sigma_k)^\top$.

\subsubsection{Model parameters}

To select  the five internal parameters of the system 
in a physically meaningful way, we related them to the 
five external parameters: 
the mean  time scale $\bar\tau_x$ of the process $x_k$,
the  time scales $\tau_F$ and $\tau_\Sigma$ of the processes $F_k$ and $\Sigma_k$,
the probability of the ``local instability'' $\pi=\Pr(|F_k| > 1)$, and
the variability in the system-noise variance, which we quantify by $\sd\Sigma_k$, the 
standard deviation of $\Sigma_k$.
We specified the external parameters and then calculated the internal ones;
we omit the respective elementary formulas.

\subsection{The ``default'' configuration of the experimental system}

\subsubsection{Model}

 In order to assign specific values to the five external parameters,
we interpreted our system, Eqs.(\ref{emdlx})--(\ref{eSigma}),
as a very rough model of the Earth atmosphere. 
Specifically, we arbitrarily postulated that one time step in our system 
corresponds to 2 hours of time in the atmosphere.
This implies that the weather-related characteristic time scale of 1 day in the atmosphere corresponds to 
the mean time scale $\bar\tau_x=12$ time steps for our process $x_k$.
This was the default value for $\bar\tau_x$ in the experiments described below.
Further, for the ``structural'' time series $F_k$ and $\Sigma_k$, 
we specified somewhat longer time scales, $\tau_F=\tau_\Sigma=1.5 \bar\tau_x$.
Next, the default value of $\pi$ was selected to be equal to 0.05 and
$\sd\Sigma_k$ equal to 0.5---these two values gave rise to reasonable 
variability in the system.
We also examined effects of deviations of $\pi$ and $\sd\Sigma_k$   from their default values, 
as described  below. The sensitivity of our results to the other parameters of the model appeared to be low.

\subsubsection{Observations}
\label{sec_obs}

We generated observations by applying Eq.(\ref{yHx2}) every time step with $H_k=1$ and
$\eta_k \sim {\cal N}(0, {R})$ (so that the observation error variance $R_k=R$
is constant in time).
To select the default value of $R$, we specified the default  
ratio $B/R$. In meteorology, for most observations,
this forecast error to observation error ratio is about 1, 
but only a fraction of all system's degrees
of freedom is observed. In our scalar system, the only degree of freedom
is observed, so, to mimic the sparsity of meteorological observations, we inflated
the observational noise and so reduced the default ratio $B/R$ to be equal to $0.1$.
This appeared to roughly correspond to the default $\sqrt{R}=9$.
We also examined the effect of varying $R$: from the well observed case with $B/R \simeq 10$
to the poorly observed case with $B/R \simeq 0.01$.

\subsubsection{Ensemble size}

In real-world atmospheric applications $N$ is usually several tens or hundreds
whilst the dimensionality of the system $n$ is up to billions.
In our system $n=1$, so we chose $N$ to vary from 2 to 10 with the default
value of $N=5$.

\subsubsection{Version and parameters of the HBEF}

By default, the simplest version of the HBEF was used, see section \ref{sec_simpl_an}.
To complete the specification of the default HBEF, it remained to assign values to the three
sharpness parameters $\chi$, $\phi$, and $\theta$, which was done by manual tuning. The default 
respective values were $\chi=5$, $\phi=30$, and $\theta=2$.

\subsubsection{Other parameters of the experimental setup}

In  the EnKF,  the tuned variance inflation factor was  1.005.
In the HEnKF, the  best  sharpness parameter was found to be $\theta=10$.
If not stated otherwise, the below statistics were computed with the length of 
the time series (the number of assimilation cycles) equal to $2\cdot 10^5$.

\subsection{Estimation of  the true prior variances ${B}_k$ and signal variances ${V}_k$}
\label{sec_est_tru_var}

For an in-depth exploration of the HBEF's secondary filter, knowledge of  
the {\em true} forecast error variance ${B}_k$ is very welcome, just like exploring the behavior
of a primary filter is facilitated if one has access to the truth $x_k$.
In this section, we show that our experimental methodology enables the assessment of
the true  ${B}_k$ as accurately as needed.

We start by noting that each filter produces estimates of its own forecast error (co)variances $B_k$.
By construction, the (exact) KF produces forecast error variances 
that coincide with the true ${B}_k$. All approximate filters (including those considered 
in this study) can produce only {\em estimates} of the ${B}_k$, \eg the HBEF produces
the posterior estimate $B^a_k$, see Eq.(\ref{BaDet}).
It is worth stressing that  $B_k$ produced by the KF cannot be used as a proxy to the true $B_k$
of any other filter because the error (co)variances are filter specific. 
The true ${B}_k$ for each filter and each $k$ can be assessed as follows.

Recall that ${B}_{k}$ is the conditional (given all assimilated data)
forecast error variance. Two aspects are important  for us here.
(i) ${B}_{k}$ is the {\em forecast error variance}; this suggests that it can be assessed 
by averaging squared errors of the deterministic forecast, $(x^f_k - x_k)^2$.
(ii) ${B}_{k}$ is the {\em conditional} error variance; this means that ${B}_{k}$ 
depends on all assimilated so far observations, so 
in order to assess the true ${B}_{k}$,  one has to 
perform the averaging of squared errors only for those trajectories of the truth and those 
observation errors that give rise to exactly (or even approximately) the same observations
${B}_{k}$ is conditioned upon. 
This is a computationally
unfeasible task even for a one-variable model. 
But the assessment of the {\em unconditional}
forecast error covariance matrix $\underline{B}_{k}$ is feasible and 
parallels the estimation of the true variance ${V}_k$ outlined in section \ref{sec_truth_mdl_compar}.

Specifically, we performed  $L$ independent assimilation runs, in  which
 the sequences of $F_{k}$ and $\sigma_{k}$ (as well as the sequence of the observation operators)
 were the same
(thus preserving the specificity of each time instance), 
whereas the sequences of ${\varepsilon}_k$, 
${\eta}_k$, and the random sources in the filters related to the generation of the analysis 
ensembles were simulated in each run randomly and independently from the other runs. 
Then we used the mean squared forecast error as a proxy to the true $\underline{B}_{k}$: 
\begin {equation}
\label{hatB}
 \underline{\hat B}_{k} = \langle (x^f_k - {x}_k)^2 \rangle,
\end {equation}
where the angle brackets $\langle . \rangle$ denote averaging over the $L$ runs.
In our experiments  $L=500$.

As noted in remark \ref{remarks_kf_2} in section \ref{sec_remarks_kf}, the KF's conditional 
${B}_{k}$ does not depend on the assimilated observations at all and thus coincides
with the unconditional $\underline{B}_{k}$. This is true for any 
non-adaptive EnKF, the HEnKF, and the simplest version of the HBEF as well. 
But for the HBEF with the Monte Carlo based analysis, where there is feedback from observations to the
covariances, this is not exactly the case. However, as we discussed in section \ref{sec_simpl_an},
the influence of observations on the posterior estimates of ${P}_{k}$ and ${Q}_{k}$ 
(and thus ${B}_{k}$) is relatively weak,
so we used $\underline{B}_{k}$ as a proxy to  ${B}_k$ for 
the HBEF with the Monte Carlo based analysis as well.
To simplify the notation, 
we do not distinguish (for any filter in question) between the true conditional variance ${B}_k$, 
the true unconditional variance $\underline{B}_k$, and the proxy 
$\underline{\hat B}_{k}$.

Thus, for any time $k$, we had at our disposal  the  variance of the truth ${V}_k$ and each filter's true
forecast error variance ${B}_k$.

\subsubsection{Remarks}
\label{sec_remarks_est_tru_var}

\begin{enumerate}
\item
Our approach here is similar to that proposed in \cite{Bishop}. 
The difference is that in \cite{Bishop} the truth is deterministic (so that ${V}_k$ cannot be assessed)
and the forecast model is stochastic, whereas our model assumes that the truth is stochastic
whilst the forecast model is deterministic.

\item
In order to avoid confusion with the filters' internal {\em estimates} of $B_k$ (\eg  $B^a_k$), 
we use the terms {\em assessment} or {\em proxy} to refer to $\underline{\hat B}_{k}$,
which externally evaluates the actual performance of the filter using the access to the truth.

\item
All  numerical experiments presented in this paper were carried out 
with one and the same arbitrarily selected realization
of the structural time series $F_k$  and $\sigma_k$, 
so that for any $k$, the signal variance ${V}_k$
is the same for all  plots below. This holds also for any filter's true
forecast error variance ${B}_k$, facilitating  comparison of the different plots.

\end{enumerate}

\subsection{Model's behavior}

\begin{figure}
\begin{center}
   { \scalebox{0.82}{ \includegraphics{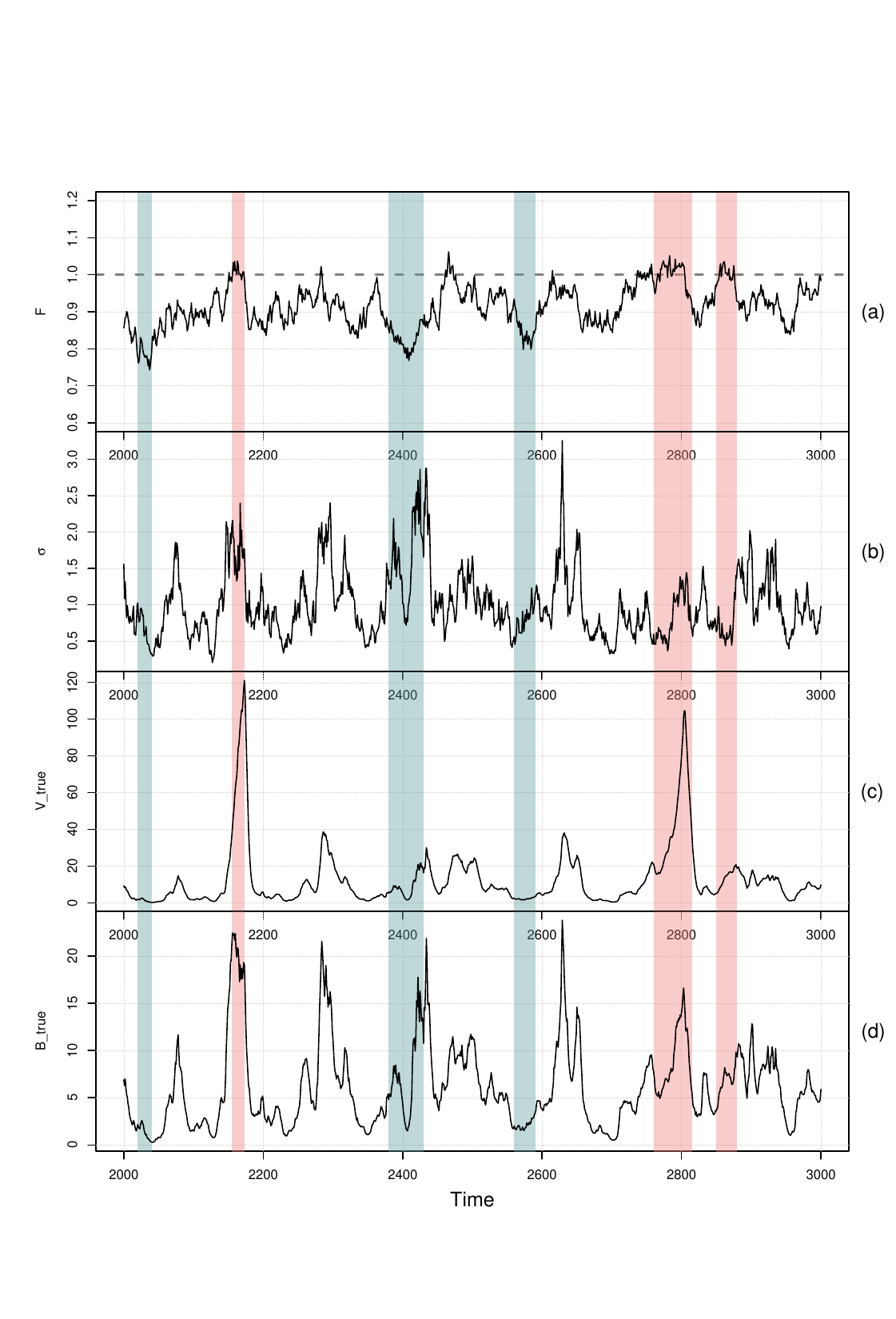}}
       }
\end{center}
  \caption{Typical time series of: 
(a) The forecast operator $F_k$, 
(b) The model error standard deviation $\sigma_{k}$, 
(c) The variance of the truth ${V}_k$, and
(d) The true background error variance ${B}_k$ for the HBEF.
The light gray (pink in the web version of the article) vertical stripes indicate events when $F_k>1$.
The dark  gray (blue in the web version of the article) vertical stripes indicate events when $|F_k|$ was relatively low.}
\label{Fig_time}
\end{figure}

Figure \ref{Fig_time} displays typical time series segments of
$F_k$ and $\sigma_k$, as well as of the true signal variance ${V}_k$ and 
the HBEF's true forecast error variance ${B}_k$. 
One can see that the 
variance ${V}_k$ of the signal $x_k$ can vary in  time  by as much as some two orders of magnitude, 
so  the process $x_k$ was significantly non-stationary, 
as it is the case, say, in meteorology.
One can also observe that 
the system-noise standard deviation $\sigma_k$ was correlated
 with both ${V}_k$ and ${B}_k$ (which is not surprising). 
Correlation between $F_k$ and both ${V}_k$ and ${B}_k$ was
also positive but lower.
Both ${V}_k$ and ${B}_k$ tended to be high when both $|F_k|$ and $\sigma_k$ are high 
(low-predictability events),
and low when both $|F_k|$ and $\sigma_k$ are low (high-predictability regimes).
In general, the model behaved as expected.

\subsection{Verifying the conditional Gaussianity of the state given $(x^f,{B})$}
\label{sec_verif_Gau}

\begin{figure}
\begin{center}
   { \scalebox{0.55}{\includegraphics{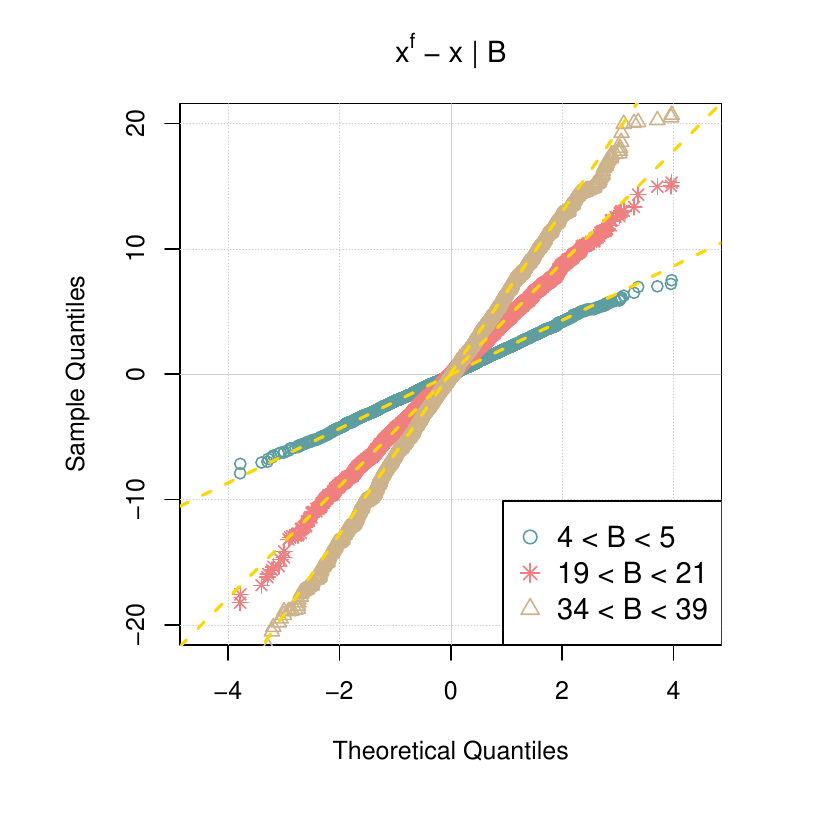}}
     \scalebox{0.55}{\includegraphics{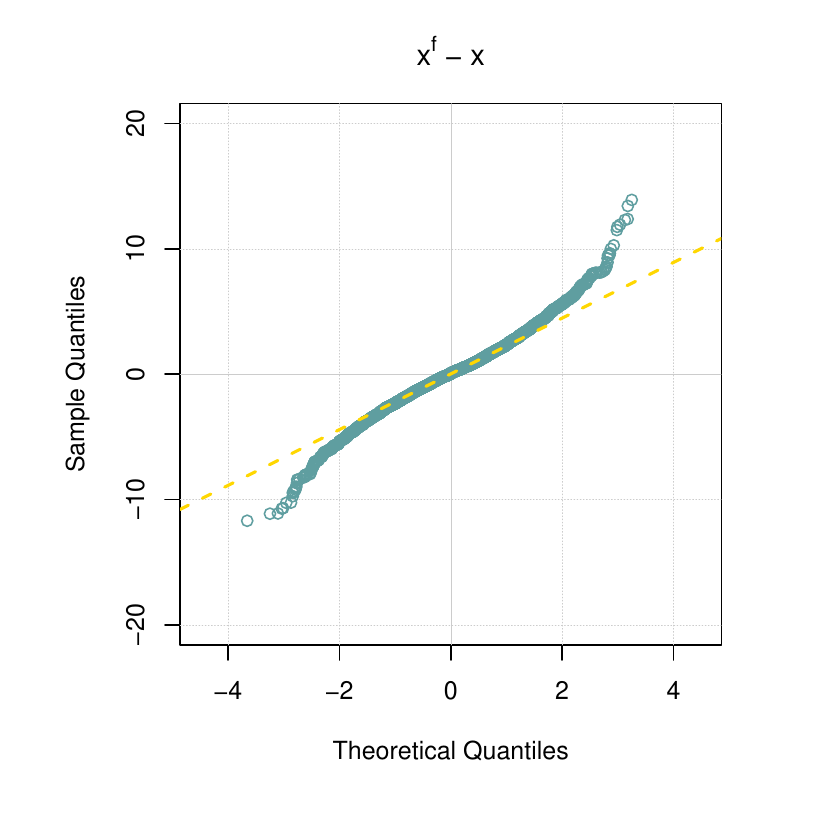} } }
\end{center}
  \caption{The Gaussian $q$-$q$ plots for the conditional pdf  $p(x_k^f - x_k|{B}_k)$ (left) and the
   unconditional pdf $p(x_k^f - x_k)$ (right). 
   In the left panel, the three curves 
   correspond to the three intervals of ${B}_k$ indicated in the legend.}
\label{Fig_xB}
\end{figure}

From the equation
\begin {equation}
\label{pfx2}
{x} | x^f,{P}, {Q} \sim {\cal N}({x}^f, B={P}+{Q}),
\end {equation}
it is obvious that $x_k | x_k^f,B_k$ 
is Gaussian if and only if so is  $x_k - x_k^f| B_k$.
With the  true ${x}_k$  and ${B}_k$ in hand, we were able to  verify if indeed 
${x}_{k} - {x}^f_{k} \, | \, {B}_{k} \sim {\cal N}(0, {B}_{k})$.
Fig.\ref{Fig_xB}(left) presents the respective $q$-$q$ (quantile-quantile) plots.
(Note that for a Gaussian density, the $q$-$q$ plot is a straight line, with the slope proportional to 
the standard deviation of the empirical distribution.)

One can see that $p(x_k- x_k^f|{B}_k)$ can indeed be very well approximated 
by a Gaussian density for low, medium, and high values of ${B}_k$ 
(the three curves in Fig.\ref{Fig_xB}(left)).
In contrast, the {\em unconditional} density $p(x_k-x_k^f)$ is 
significantly non-Gaussian with heavy tails, see Fig.\ref{Fig_xB}(right). 
So, the conditional Gaussianity of the state's prior distribution
is confirmed in our numerical experiments.

\subsection{The forecast ensemble members are {\bf not} drawn from the same distribution as the truth}
\label{sec_Pobs}

Here, we explore the actual probability distribution of the forecast ensemble members at any given time $k$.
We demonstrate that for both the EnKF and the HBEF, the variance of this distribution
is often substantially biased with respect to the respective true error variance.

We start by stating that in a single data assimilation run, we cannot find out from which
(continuous) probability distribution
the forecast ensemble members at time $k$ are drawn (because the ensemble size is small, see
assumption  \ref{list_setup_N}). 
But, following section \ref{sec_est_tru_var}, for each filter,
we had at our disposal a number of assimilation runs
that share the sequence of  ${B}_k$.
Then, {\bf if} in each assimilation run, 
the forecast ensemble members were drawn from the  distribution with the 
variance ${B}_k$ (the ``null hypothesis''),
we would have $\Ex S_k = {B}_k$, where 
$S_k$ is the ensemble (sample) variance and 
the expectation is over the population of 
independent assimilation runs. 
To check if this latter equality actually holds, 
we  estimated $\Ex S_k$ as the sample mean $\langle S_k \rangle$ for each $k$ separately
using the sample of $L$  assimilation runs.

\begin{figure}
\begin{center}
   { \scalebox{0.85}{ \includegraphics{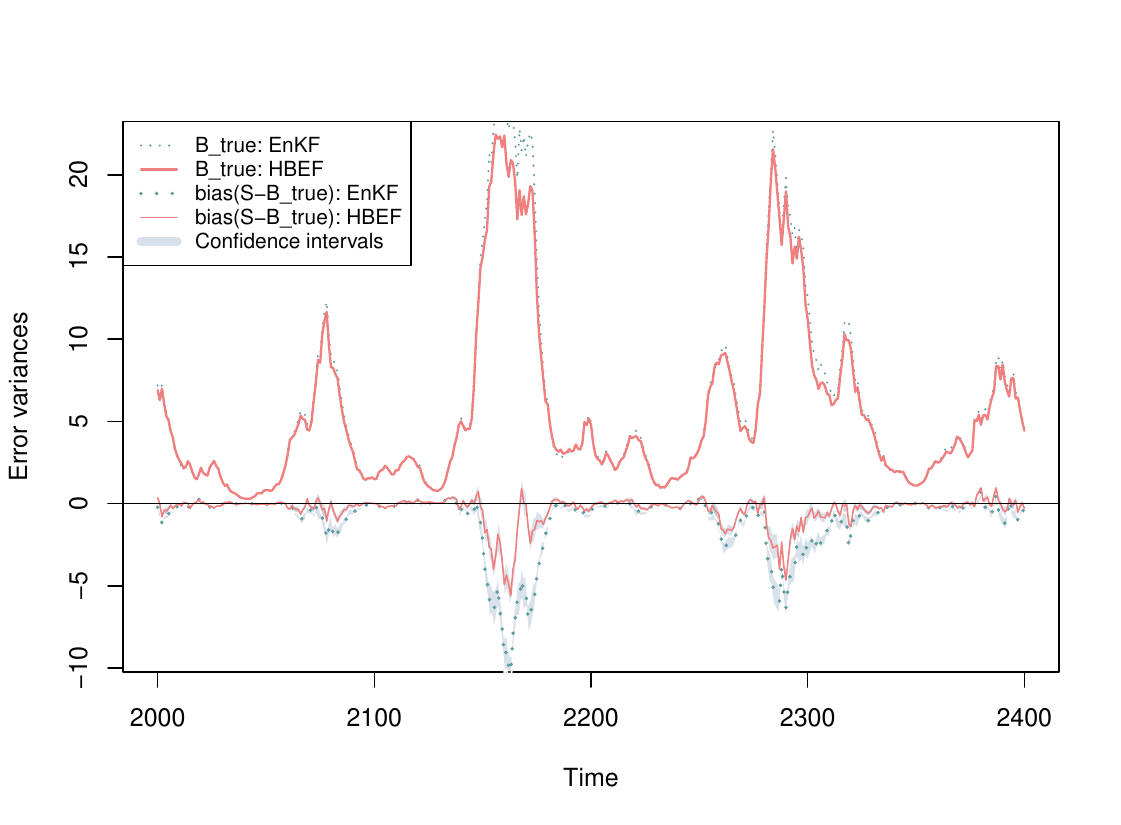}}
       }
\end{center}
  \caption{The two lower curves: 
  biases  in the forecast ensemble variances
  (with the 95\% confidence intervals) for the EnKF  and the HBEF.
  The two upper curves: the respective true error variances ${B}_k$.
  }
\label{Fig_S_bias}
\end{figure}

The resulting time series of the {\em biases} $\langle S_k \rangle - {B}_k$ for the EnKF and the HBEF
are displayed in Fig.\ref{Fig_S_bias}
(the two lower curves) along with their respective 95\% bootstrap confidence intervals.
The true error variances themselves ${B}_k$ are also shown in Fig.\ref{Fig_S_bias}
(the two upper curves) to give an impression of the relative magnitude of the biases in $\langle S_k \rangle$.

One can see that that the biases in the ensemble variances were significantly non-zero 
when the true $B_k$ were relatively large.
For the EnKF, the deviation of    $\langle S_k \rangle$
from ${B}_k$  sometimes reached 50\% of  ${B}_k$.
For the HBEF, the biases were less but still significant.
In the small forecast error regimes, the biases became insignificant.
It is also interesting to notice that the  large biases were mostly negative implying that the filters
were under-dispersive (despite the tuned variance inflation  in the EnKF).
Over a longer time window of $10^4$ time steps, the confidence interval did {\em not} contain zero
(\ie the bias was significantly non-zero) 78\% of time for the EnKF and 62\% of time for the HBEF.
 
Thus, we have to reject the null hypothesis and admit that forecast ensemble members are 
often taken from a distribution which is significantly different from the true one.
This has two implications.
First, the uncertainty in the sample covariances is not only due to the sampling noise but also due
to an accumulated in time systematic error component.
Second, the biases in the sample covariances warrant the introduction 
of the actual predictability ensemble covariance matrix $\boldsymbol\Pi_k$ 
that differs from the true covariance matrix ${\bf P}_k$  (see section \ref{sec_pe_lik}).
 
The above results are worth comparing with those of Bishop and Satterfield \cite{Bishop}, 
who found insignificant biases in the ensemble variances, see their Fig.2. 
One dissimilarity between their and our experiments 
was that an ensemble transform version of the EnKF was used in \cite{Bishop}.
We employed the ensemble transform technique for both the EnKF and the HBEF
and found that this led to some improvements but did not remove
the biases in $S_k$ (not shown).
A plausible reason for the difference in the conclusions is that the system in \cite{Bishop}
was much better observed than ours (they used $R$ which was much less than the  mean $V_k$,
whereas in our study $R$ was several times larger than the mean $V_k$).

\subsection{Verifying the primary filters}
\label{sec_perf_prim}

Here, we examine the accuracy of the state estimates for the HBEF and the other filters
(the Var, the EnKF, and the HEnKF).
In the below figures, we display their analysis RMSEs with the reference-KF analysis RMSEs subtracted.

\begin{figure}
  \begin{subfigure}[b]{0.5\linewidth}
    \centering
    \includegraphics[width=0.95\linewidth]{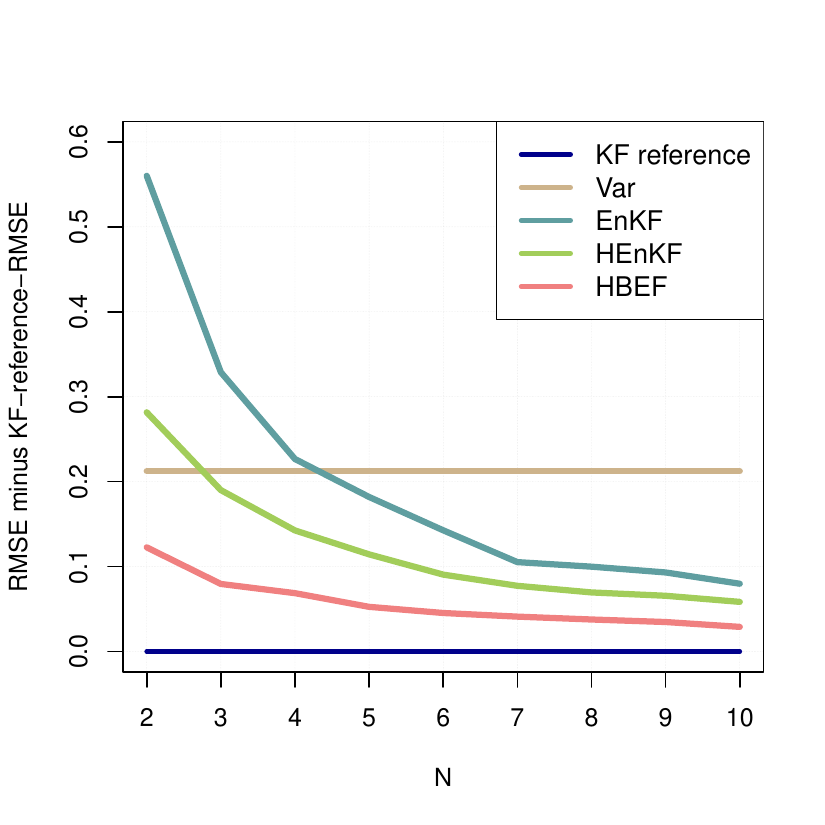} 
    \label{Fig_perf:a} 
  \end{subfigure}
  \begin{subfigure}[b]{0.5\linewidth}
    \centering
    \includegraphics[width=0.95\linewidth]{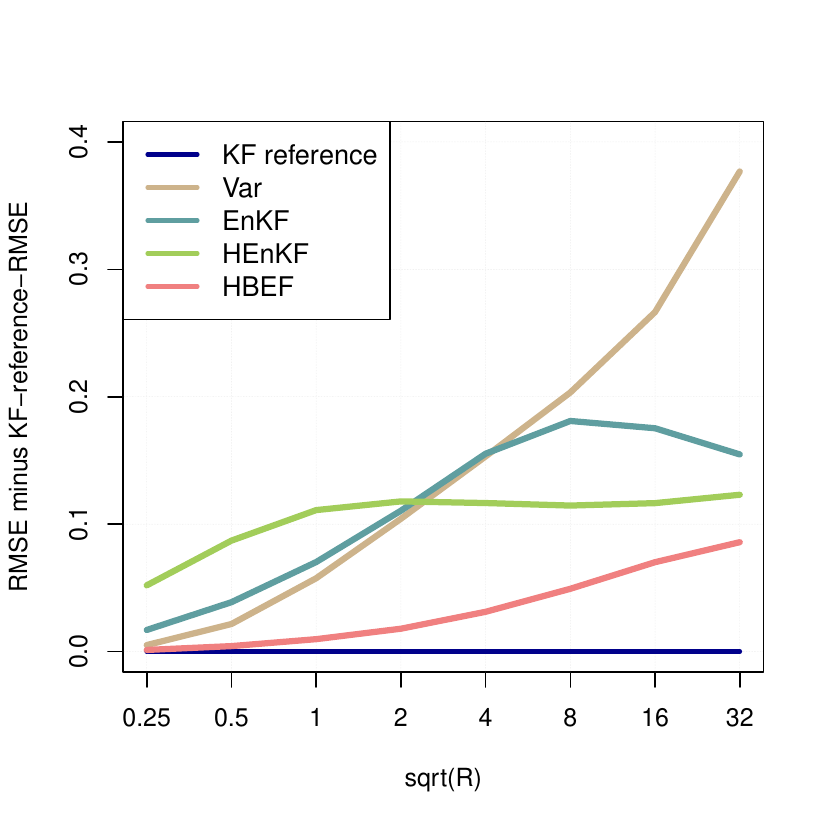} 
    \label{Fig_perf:b} 
  \end{subfigure} 
  \begin{subfigure}[b]{0.5\linewidth}
    \centering
    \includegraphics[width=0.95\linewidth]{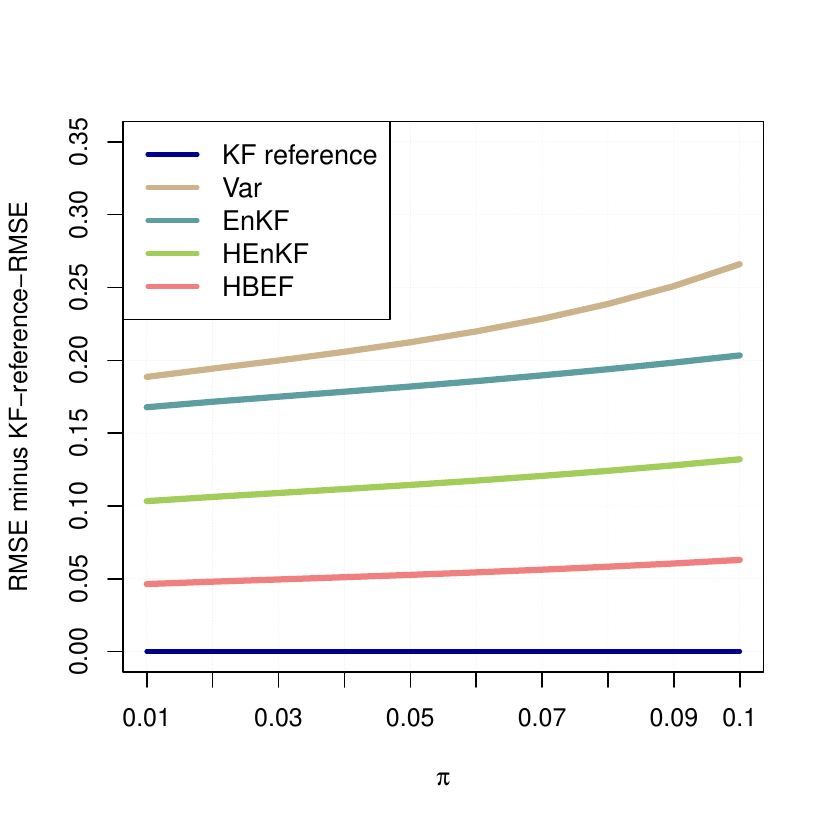} 
    \label{Fig_perf:c} 
  \end{subfigure}
  \begin{subfigure}[b]{0.5\linewidth}
    \centering
    \includegraphics[width=0.95\linewidth]{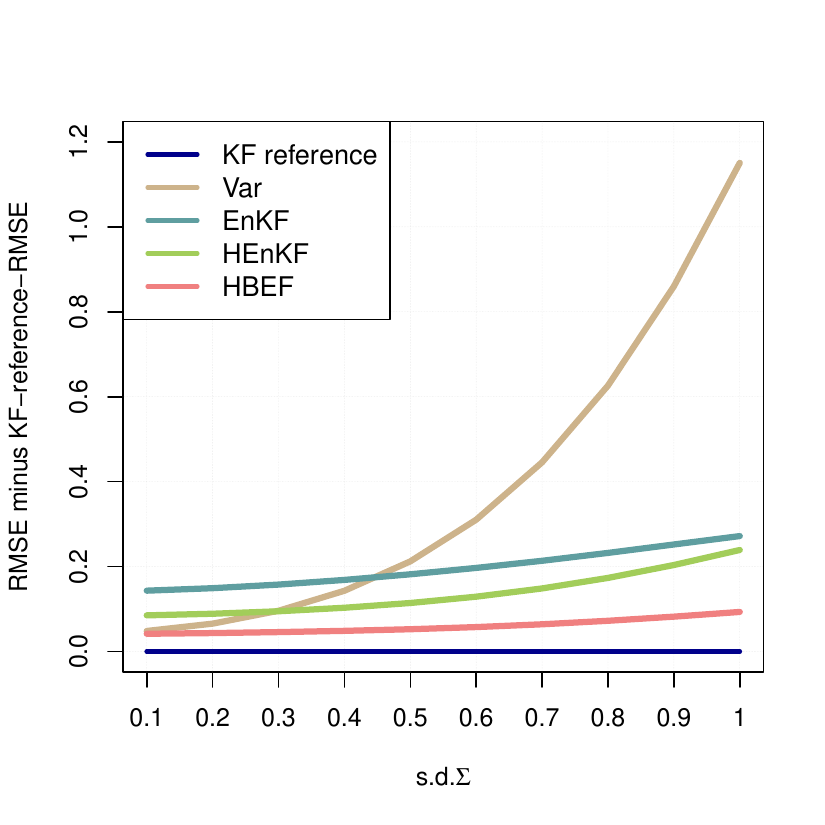} 
    \label{Fig_perf:d} 
  \end{subfigure} 
  \caption{The filters' analysis RMSEs of the state (with the reference-KF analysis RMSE subtracted) as functions of
          the ensemble size $N$ (top, left), 
          the observation error standard deviation $\sqrt{R}$ (top, right), 
          the degree of the system's intermittent instability $\pi$ (bottom, left), 
          the variability in the model error  standard deviation $\sd\Sigma$ (bottom, right).}
  \label{Fig_perf} 
\end{figure}

Figure \ref{Fig_perf}(top, left) shows the RMSEs as functions of the ensemble size $N$. 
One can see that the HBEF was by far the best filter.
For small $N < 3$, the Var became more competitive than the EnKF and the HEnKF, but still worse than the HBEF. 

Figure \ref{Fig_perf}(top, right)  shows the  RMSEs as functions of $\sqrt{R}$.
Again, the HBEF performed the best. 
Its relative superiority was especially substantial for the smaller values of  $\sqrt{R}$.
This can be explained by the prevalence of $Q$ 
(which is more rigorously treated in the HBEF) over $P$ (which is only sub-optimally treated  in the HBEF)
in this regime.

Figure \ref{Fig_perf}(bottom, left) shows the RMSEs as functions of  $\pi=\Pr(|F_k| > 1)$. 
One can see that the HBEF was uniformly and significantly better than the other filters.
Note that all the filters gradually deteriorate \wrt the reference KF as the system becomes less
stable (\ie as $\pi$ grows), which is meaningful because errors grow faster in a less stable system.

Figure \ref{Fig_perf}(bottom, right) displays  the RMSEs as functions of the degree of 
intermittency in the model error variance quantified by $\sd\Sigma$.
We see that the HBEF was still uniformly and substantially better than the EnKF and the  HEnKF. 
For the smallest values of $\sd\Sigma$, the Var became superior  to the EnKF and the HEnKF and only slightly
worse than the HBEF.
The fact that the Var worked relatively better for the small $\sd\Sigma$
can be explained by noting that 
in this regime, when the variability in $Q$ was low, 
the forecast error statistics were less variable
and so the constant Var's $\bar B$ was relatively more suitable. 

Thus, in terms of the analysis RMSEs,
 the HBEF  demonstrated its overall superiority over the competing EnKF, HEnKF, and Var filters.

\subsection{Verifying the secondary filters}
\label{sec_ver_scnd}

Recall that the HBEF's secondary sub-filter produces the  posterior estimate
 ${B}^a_k={P}^a_k+{Q}^a_k$  of its  true forecast error variance ${B}_k$.
The HEnKF yields its ${B}^a_k$  as described in item (iii) in section \ref{sec_henkf}.
The Var uses the constant  $\bar B$ as an estimate of   ${B}_k$, 
so we associate $\bar B$ with its ${B}_k^a$.
Similarly, we identify the EnKF's inflated ensemble variance ${S}_k$ with its ${B}^a_k$.

\begin{figure}
\begin{center}
   { \scalebox{0.70}{ \includegraphics{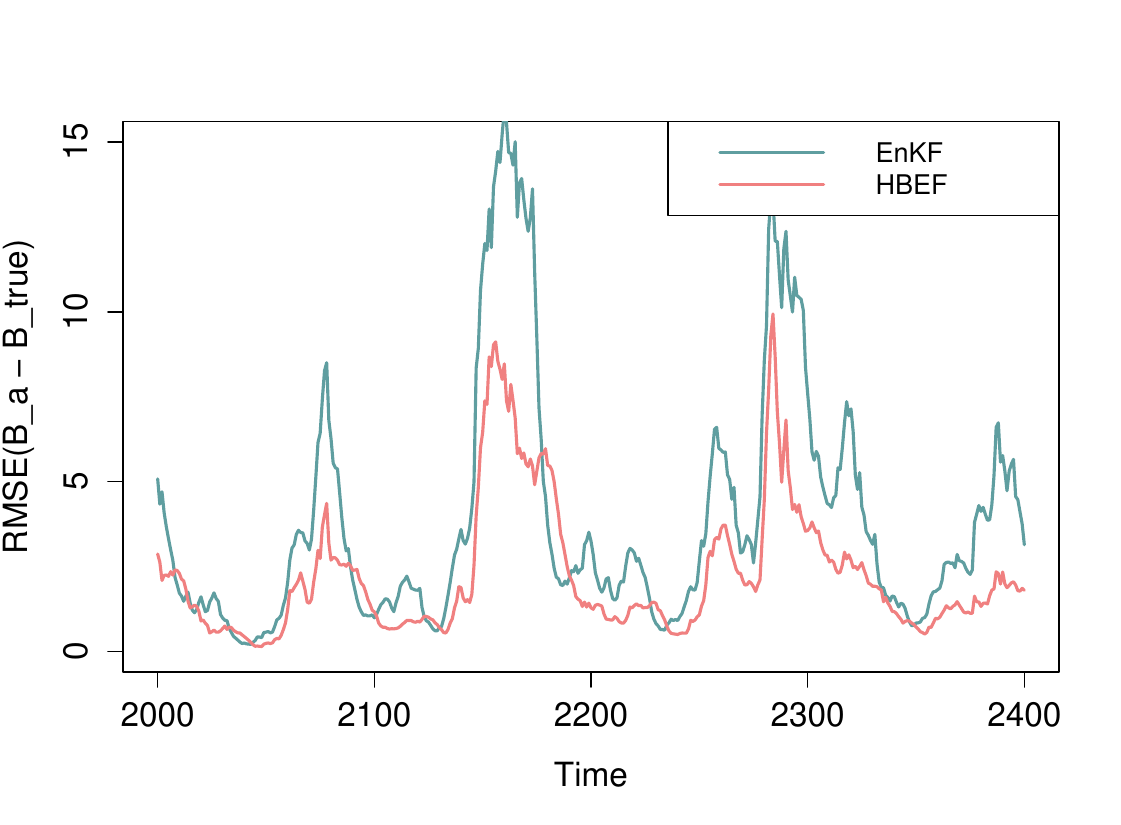}}
       }
\end{center}
  \caption{RMSEs  in $B^a_k$ produced by the EnKF and the HBEF.
  }
\label{Fig_RMSE_B}
\end{figure}

In this section, we  examine the errors   ${B}^a_k - B_k$, with the filter specific ${B}_k$
assessed following section  \ref{sec_est_tru_var}.
Having the true $B_k$ for each filter, we computed 
the RMSE in its $B^a_k$ estimates using  averaging over the $L$
 independent assimilation runs as 
$\Delta_k =\sqrt{ \langle ({B}^a_k - {B}_k)^2  \rangle }$.
The resulting $\Delta_k$ for the HBEF and the EnKF are depicted in Fig.\ref{Fig_RMSE_B}, 
where the almost uniform and substantial superiority of the HBEF is evident.

Having square averaged $\Delta_k$ over time, we obtained the time mean RMSEs in ${B}^a_k$.
In a similar way we computed the biases in ${B}^a_k$.
The results of an experiment with $10^4$ time steps are collected in Table \ref{Tab_accB},
where it is seen that the HBEF was much more accurate in estimating its ${B}_k$ than
the Var, the EnKF, and the HEnKF in estimating their respective true forecast error variances.

\begin{table}[ht]
\caption{Accuracy of the filters' estimates of their own forecast error variance $B_k$}

\begin{center}
\begin{tabular}{lccc}

\hline
Filter & Error bias & RMSE  & Mean true $B$\\
	  & $\mean({{B}^a_k -{B}_k})$  & $\rms ({B}^a_k - {B}_k)$  & $\mean({{B}_k})$ \\
\hline
Var                & -0.9   & 6.5  &   7.6  \\
EnKF              & -1.4   & 6.2  &   7.5  \\
HEnKF           & -1.8   & 4.4  &   7.2  \\
HBEF             & -0.5   & 3.2  &   7.0  \\
\hline
\end{tabular}
\end{center}
\label{Tab_accB}
\end{table}
%

\subsection{Role of feedback from observations to forecast error covariances}
\label{sec_feedb}

The  HBEF with the Monte Carlo based analysis (section \ref{sec_MC_an})
provides an optimized way to utilize observations
in updating $P$ and $Q$.
In the default setup, this capability did not lead to any improvement in the performance scores (not shown), 
but it became significant when the filter's 
model error variance was {\em misspecified}. 

\begin{figure}
\begin{center}
   { \scalebox{0.70}{ \includegraphics{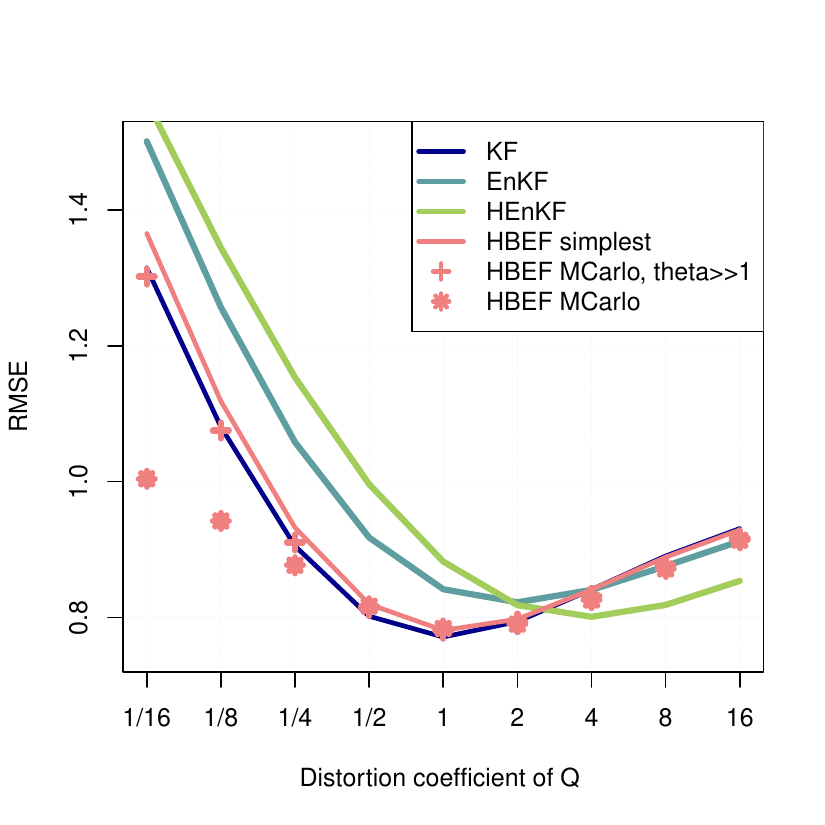}}
       }
\end{center}
  \caption{Analysis  RMSEs  of the state  for the filters which used the  
  wrong   $Q$. The latter was specified to be the true $Q$ multiplied by the 
  distortion coefficient $q_{distort}$.
    }
\label{Fig_RMSE_Q_distort}
\end{figure}

Specifically, we let all the filters  (including the KF)
``assume'' that the model error variance $Q_k$ equals the true one multiplied by 
the distortion coefficient $q_{distort}$.
For several values of ${q_{distort}}$ in the  range from $1/16$ to $16$, we computed the RMSEs of the analyses of the state 
for all filters and plotted the results in Fig.\ref{Fig_RMSE_Q_distort}.
In the HBEF with the Monte Carlo based analysis, the size of the Monte Carlo sample was $M=100$, see Eqs.(\ref{PQaDetMC})--(\ref{xaDetMC}).

To make the effect more pronounced, the observation error standard deviation was reduced to $\sqrt{R}=1$.

From Fig.\ref{Fig_RMSE_Q_distort} one can see that the overall 
performance of the HBEF with the Monte Carlo based analysis was  better than 
the performances of the other filters, 
including, we emphasize, the (now, inexact) KF. 
The observations-to-covariances feedback present in the Monte Carlo based HBEF 
(and absent in the other filters)
appeared especially useful for $q_{distort} < 1$. 
The improvement was bigger for $q_{distort} < 1$ than for $q_{distort} > 1$ because an underestimation of the
forecast error covariances is potentially more problematic for any filter.
Indeed, the overconfidence in the forecast leads to an underuse of observations
and in extreme cases can even lead to filter divergence. This is why the settings
with $q_{distort} < 1$ left more room for improvement, particularly due to the 
feedback from observations to the covariances.

Another interesting conclusion can be drawn from comparing the Monte Carlo based version of the HBEF
with the optimally tuned parameter $\theta$ (asterisks in Fig.\ref{Fig_RMSE_Q_distort})
and the same version  of the HBEF but with $\theta=\infty$ (crosses).
Recall that $\theta$ controls the difference between the variance $\Pi$ of the distribution
of  the predictability ensemble members  and the true variance $P$. In the setting with $\theta=\infty$,
the HBEF ``assumes'' that $\Pi=P$. Figure \ref{Fig_RMSE_Q_distort} clearly shows that 
it was indeed beneficial to get away from the traditional assumption  $\Pi=P$. 
This again justifies our suggestion (see section \ref{sec_pe_lik}) 
to allow the ensemble distribution to be different from the true one.

\section {Discussion}

\subsection{Comparison with other approaches}

The HBEF has two immediate predecessors,  the HEnKF  \cite{Myrseth} and 
the EnKF-N \cite{Bocquet,Bocquet2015}. The HBEF differs from the HEnKF in the following aspects.
First, in the HBEF we treat ${\bf Q}$ and ${\bf P}$ separately instead of using the
total background error covariance matrix ${\bf B}$.
Second, the HBEF's forecast step is based on the persistence forecasts for the posterior point estimates of 
${\bf Q}$ and ${\bf P}$ instead of that for the analysis error covariance matrix.
These two improvements have led to the substantially better performance of the HBEF as compared to the HEnKF.
Another difference from the HEnKF is that the Monte Carlo based HBEF permits observations to influence ${\bf Q}$ and ${\bf P}$.
Experimentally, this latter feature appeared to be beneficial only when ${\bf Q}$ was significantly misspecified, though.

As compared to the EnKF-N, which integrates ${\bf B}$ out of the prior distribution, 
the HBEF explicitly updates the covariance matrices. This introduces 
memory in the covariances, which, as we have seen in the numerical experiments, can be beneficial. 

In contrast to both the HEnKF  and the EnKF-N,  the HBEF in its present formulation 
does not treat the uncertainty in the prior mean state vector (this may be worth exploring in the future).
But the HBEF systematically treats the uncertainty in ${\bf Q}$, which was assumed to be known
in \cite{Myrseth} and equal to zero in \cite{Bocquet,Bocquet2012,Bocquet2015}.

\subsection{Restrictions of the proposed technique}

First, the HBEF heavily relies on the {\em conditional} Gaussian prior distribution of the state.
It is this assumption  that greatly simplifies the analysis algorithm, but in a nonlinear context, it becomes an approximation,
whose validity is to be verified. 

Second, the HBEF makes use of the inverse Wishart prior distribution for 
the covariance matrices.
There is no justification for this hypothesis other than partial analytical tractability
of the resulting analysis equations, so other choices can be explored.

\subsection{Practical applications}

In order to apply the proposed technique to real-world high-dimensional problems,
simplifications are needed because the $n \times n$ covariance matrices will be too large to be stored
and handled. The computational burden can be reduced in different ways. 
Here is one of them.
First, let the covariances to be defined on a coarse grid.
Second, localize (taper) the covariances and store only non-zero covariance matrix entries.
Third, use the simplest version of the HBEF.

Another possibility is to fit a parametric covariance model to current covariances and 
impose persistence for the {\em parameters} of the  model. In this case, the simplest version
of the HBEF would become close to practical ensemble variational schemes, but with
climatological covariances replaced by evolving recent-past-data based covariances.

In high dimensions, the persistence forecast for the covariances seems to be worth improving.
Specifically, one may wish to somehow spatially smooth 
${\bf P}^a_{k-1}$ and ${\bf Q}^a_{k-1}$ in Eq.(\ref{fcPQ})---because it is meaningful that smaller scales in 
${\bf P}^a_{k-1}$ and ${\bf Q}^a_{k-1}$ 
have less chance  to survive until the next assimilation cycle than larger scales.
Another way to improve the empirical forecast of the covariance matrices
is to introduce a kind of ``regression to the mean'' making use of the time mean
covariances. This would imply that the HBEF  would cover not only EnKF but also ensemble variational hybrids
as a special case.

The ultimate goal with the HBEF
will be to obtain  effective covariance  regularization as a by-product of the hierarchical analysis scheme
without using any ad-hoc device
(as it was proposed for the EnKF-N in \cite{Bocquet} and partially tested in \cite{Bocquet2015}).

\section {Conclusions}

The progress made in this study can be summarized as follows.
\begin{itemize}
\item
We  have acknowledged that in most applications, 
the EnKF works with: 
 (i) the explicitly unknown and variable model error covariance matrix ${\bf Q}_k$,
 (ii) the  partially known (through ensemble covariances) 
  background error covariance matrix.
Under these explicit restrictions, we have proposed a new Hierarchical Bayes Ensemble  Filter (HBEF)
that optimizes the use of observational and 
ensemble data by treating ${\bf Q}_k$ and the predictability covariance matrix 
${\bf P}_k$ as random matrices to be estimated in the analysis along with the state.
The ensemble members are treated in the HBEF as generalized observations on the covariance matrices.
\item
With the new HBEF filter, in the course of filtering,
the prior and posterior distributions of the state 
 remain conditionally (given ${\bf P}_k, {\bf Q}_k$) Gaussian 
provided that: 
(i) it is so at the start of the filtering, 
(ii) observation errors are Gaussian, 
(iii) the dynamics and the observation operators are linear, and 
(iv) model errors are conditionally Gaussian given ${\bf Q}_k$. 
Unconditionally, the prior and posterior distributions of the state are non-Gaussian.
\item
The HBEF is tested with a new one-variable doubly stochastic model of truth.
The model has the advantage of providing the means to assess the instantaneous variance of the truth
and the true filter's error variances.
The HBEF is found superior  the EnKF and the HEnKF \cite{Myrseth}
under most regimes of the system, most data assimilation setups, and in terms of performance
of both  primary and  secondary filters.

\item
The availability of the true error variances has permitted us to experimentally  prove that
the forecast ensemble variances in both the EnKF and the HBEF are   often significantly biased 
with respect to the true variances.
\item
It is shown that the HBEF's feedback from observations to the covariances can be beneficial.
\item
The simplest version of the HBEF is designed to be affordable 
for practical high-dimensional applications on existing computers.

\end{itemize}
%

\section {Acknowledgments}

The authors are very grateful to the two anonymous reviewers, whose valuable comments helped to  
significantly improve the manuscript.

\section*{References}

\bibliography{mybibfile}

\let\thefigureSAVED\thefigure

\appendix 
\section{Inverse Wishart distribution}
\label{app_W}

In Bayesian statistics, the inverse Wishart distribution (\eg\cite{gupta, AndersonT, Gelman}) is the standard choice for the prior
distribution of a random covariance matrix, 
because inverse Wishart is the so-called conjugate
distribution for the Gaussian likelihood,  \eg\cite{AndersonT, Gelman}.
The inverse Wishart  pdf is defined for symmetric matrices and is non-zero for positive definite ones:
\begin {equation}
\label{IW}
p({\bf Z}) \propto \frac{1} {|{\bf Z}|^{\frac{\nu + n +1}{2}} } \e^{ -\frac12 \tr ({\bf Z}^{-1} {\boldsymbol\Sigma}) },
\end {equation}
where $\nu>n+1$ is the so-called number of degrees of freedom (which controls the spread
of the distribution: the greater $\nu$, the less the spread) and ${\boldsymbol\Sigma}$ is the  
positive definite scaling matrix. 
Using the mean value  
${\bf\bar Z} = \Ex {\bf Z} =  {\boldsymbol\Sigma} / (\nu - n -1)$
instead of the scaling matrix allows us to reparametrize Eq.(\ref{IW}) as
\begin {equation}
\label{IW2}
p({\bf Z}) =  p({\bf Z} | \theta, {\bf\bar Z})  \propto \frac {1} { |{\bf Z}|^{\frac{\theta}{2} + n +1} } 
   \e^{ -\frac{\theta}{2} \tr ({\bf Z}^{-1} {\bf\bar Z}) },
\end {equation}
where we have introduced a new scale parameter $\theta=\nu - n -1 >0$,
which we call the {\em sharpness} parameter (the higher $\theta$, the narrower the density).
We symbolically write Eq.(\ref{IW2}) as
\begin {equation}
\label{IW3}
{\bf Z} \sim {\cal IW}(\theta,{\bf\bar Z}).
\end {equation}
We prefer  our parametrization $(\theta,{\bf\bar Z})$ to the common one $(\nu, \boldsymbol\Sigma)$
because ${\bf\bar Z}$ has the clear meaning of the (important) mean ${\bf Z}$ matrix.
Summarizing, the inverse Wishart pdf has two parameters: the sharpness parameter $\theta$ (a scalar)
and the mean ${\bf\bar Z}$ (a positive definite matrix).

\section{Assimilation of conditionally  Gaussian generalized  observations in an update of their covariance matrix}
\label{app_W1}

Here, we outline, following \eg \cite{Gelman}, the procedure of assimilation of independent draws from 
the distribution ${\cal N}({\bf m,Z})$, where  ${\bf m}$ is the known vector and
${\bf Z}$ the unknown random symmetric positive definite matrix,
whose prior distribution is inverse Wishart with the density specified by  Eq.(\ref{IW2}).

Let us take a draw
${\bf x}^e(i) | {\bf Z} \sim {\cal N}({\bf m,Z})$, which we  interpret as a member of  an ensemble.
Then, obviously,
\begin {equation}
\label{eLik1}
p({\bf x}^e(i) | {\bf Z}) \propto \frac {1} { |{\bf Z}|^{\frac{1}{2}} } 
   \e^{ -\frac{1}{2} ({\bf x}^e(i) - {\bf m})^\top {\bf Z}^{-1} ({\bf x}^e(i) - {\bf m}) }.
\end {equation}
We stress that Eq.(\ref{eLik1}) is nothing other than the likelihood of ${\bf Z}$
given the ensemble member ${\bf x}^e(i)$. 
Further, having the ensemble 
${\bf X}^e = ({\bf x}^e(1),\dots,{\bf x}^e(N))$ of 
$N$ independent members all taken from ${\cal N}({\bf m,Z})$, we can write down
the  respective {\em ensemble likelihood} as the product of the partial likelihoods:
\begin {equation}
\label{eLik}
p({\bf X}^e | {\bf Z}) \propto \frac {1} { |{\bf Z}|^{\frac{N}{2}} } 
   \e^{ -\frac{1}{2} \sum_{i=1}^N ({\bf x}^e(i) - {\bf m})^\top {\bf Z}^{-1} ({\bf x}^e(i) - {\bf m}) } =
                                         \frac {1} { |{\bf Z}|^{\frac{N}{2}} } 
   \e^{ -\frac{N}{2} \tr ({\bf S} {\bf Z}^{-1}) },
\end {equation}
where 
\begin {equation}
\label{S}
{\bf S} = \frac{1}{N} \sum_{i=1}^N ({\bf x}^{e}(i) - {\bf m}) \, 
                                   ({\bf x}^{e}(i) - {\bf m})^\top
\end {equation}
is the sample covariance matrix.
But having the likelihood $p({\bf X}^e | {\bf Z})$ 
means that  ${\bf X}^e$ (and its members ${\bf x}^{e}(i)$) can be regarded and treated as (generalized)
{\em observations} on ${\bf Z}$.
In particular, the ensemble can be {\em assimilated} in the standard way using the Bayes theorem.
Indeed, having the prior pdf of ${\bf Z}$, Eq.(\ref{IW2}), we obtain  the posterior
\begin {equation}
\label{Zpost}
p^a({\bf Z})  \propto p({\bf Z} | \theta, {\bf\bar Z}, {\bf X}^e)
                   \propto p({\bf Z} | \theta, {\bf\bar Z}) \cdot p({\bf X}^e |  {\bf Z}) \propto 
                              \frac {1} { |{\bf Z}|^{\frac{\theta^a}{2} + n +1} } 
                            \e^{ -\frac{\theta^a}{2} \tr ({\bf Z}^{-1} {\bf Z}^a) },
\end {equation}
where
\begin {equation}
\label{Ztheta_a}
\theta^a = \theta +N \qquad \mbox{and} \qquad 
{\bf Z}^a = \frac {\theta {\bf\bar Z} + N {\bf S}} {\theta +N}.
\end {equation}
In the  right-hand side of  Eq.(\ref{Zpost}), we recognize 
again the inverse Wishart pdf (hence its conjugacy), see Eq.(\ref{IW2}), with 
$\theta^a$ being the posterior sharpness parameter and 
${\bf Z}^a$ being
the posterior mean of ${\bf Z}$. Consequently, ${\bf Z}^a$ is the mean-square optimal point estimate
of ${\bf Z}$ given both the prior and ensemble information. So, we have optimally
assimilated the (conditionally Gaussian) ensemble data to update the 
(inverse Wishart) prior distribution of the random covariance matrix.

\section{Integral \wrt a matrix}
\label{app_mx_intgl}

For a  general $n\times n$-matrix ${\bf C}$,
the integral $\int f({\bf C}) \,\d {\bf C}$ of a scalar function  $f({\bf C})$ over the space of all matrices with real entries
 is defined as follows.
First, we {\em vectorize} ${\bf C}$, \ie build the vector $\vec{\bf C}$ of length $n^2$ that comprises all entries of ${\bf C}$.
Then, we simply identify $\int f({\bf C}) \,\d {\bf C}$ with $\int f({\bf C}) \,\d \vec{\bf C}$, that is,
with the traditional multiple (Lebesgue or Riemann) integral over the Euclidean space of dimensionality $n^2$.

The integral \wrt a symmetric positive definite matrix is defined in a similar way.
The difference from the general matrix case is that the vectorization here involves collecting in $\vec{\bf C}$ only
algebraically independent matrix entries (\eg the upper triangle of ${\bf C}$) and the multiple integral 
is over the set (the convex cone) of those $\vec{\bf C}$ that correspond to positive definite matrices.

\section{List of main symbols}
\label{app_list}

\begin{longtable}{ p{.20\textwidth}  p{.80\textwidth} }

$()^a$ &  posterior (analysis) pdf (\ie conditioned on  past and current data)   and its parameters\\
$()^f$ & prior (forecast) pdf (\ie conditioned on  past data) and its parameters \\
$\widetilde ()$ & sub-posterior pdf (\ie conditioned on past  data and current ensemble data)  and its parameters\\

$()^{fe}, ()^{ae}$ & forecast ensemble /  analysis ensemble  \\
$()^{me}$, $()^{pe}$ &  model error ensemble / predictability ensemble\\

${\bar .}$ & time mean value\\
$\langle . \rangle$ &  average over  $L$ independent realizations of the truth / assimilation runs\\

${\bf A}$  & posterior  (analysis error) covariance matrix  \\
${\bf B}$ &  prior  (forecast error) covariance matrices \\
${\bf F}$ &  forecast operator \\
${\bf H}$ &  observation operator \\
$i$ & ensemble member index\\
${\bf K}$ & Kalman gain matrix \\
$k$ & time instance index \\
$L$ & number of independent assimilation runs \\
$l({\bf B|y})$ & observation likelihood of the matrix ${\bf B}$  \\
${\bf m}^a$ & posterior mean ${\bf x}$ given ${\bf P,Q}$\\
$n$ & dimensionality of state space \\
$N$ & ensemble size \\
$p$ &  probability density function (pdf) \\
${\bf P,Q,R}$ & predictability error / model error / observation error covariance matrix\\
${\bf S}$ & sample (ensemble) covariance matrix  \\
${V}_k$ &   $\Var x_k$ \\
${\bf x}$ &  state vector,  ``truth'' \\
${\bf x}^a$ &  posterior mean vector and its approximations (deterministic analysis)  \\
${\bf x}^f$ & prior mean vector (identified in this study with the deterministic forecast) \\
${\bf x}^{..}(i), {\bf x}^{e}(i)$ &  ensemble member \\
${\bf X}$ & ensemble \\
${\bf y}$ &  observation vector \\
${\bf Y}$ &  observation and ensemble data combined \\

${\cal IW}$ & inverse Wishart distribution 
              (parametrized according to  
              \ref{app_W}) \\
${\cal N}({\bf m,B})$ &  Gaussian distribution with the mean ${\bf m}$ 
and covariance matrix ${\bf B}$ \\
$\boldsymbol\varepsilon$ & model error (system noise) vector\\
$\boldsymbol\eta$ & observation error  vector\\
$\theta,\phi,\chi$ & sharpness parameters for the inverse Wishart pdfs\\
$\pi$ & portion of time the process $F_k$ is greater than 1 in modulus \\
$\sigma$ & (time-specific) model error standard deviation \\

$\Ex$ & expectation operator \\
$\rms$, RMSE & root-mean-square value / error\\
$\sd, \Var$ & standard deviation / variance\\
$\tr$ & matrix trace \\

$\propto$ & proportionality \\
$\sim$ & has (corresponds to) the probability distribution \\
$1:k$ & concatenation from the time instance $1$ to the time instance $k$ \\

\end{longtable}




\end{document}